\documentclass[twocolumn,showpacs,preprintnumbers,amsmath,amssymb,floatfix]{revtex4}
\normalsize 
\usepackage{graphicx}% Include figure files
\usepackage{epsfig} %Works faster than the graphicx package}
\usepackage{dcolumn}% Align table columns on decimal point
\usepackage{bm}% bold math
\usepackage{epstopdf}

\begin{document}

\preprint{APS/123-QED}

\title{Measurement scheme and analysis for weak ground state hyperfine transition moments through two-pathway coherent control  }

\author{J. Choi$^1$ and D. S. Elliott$^{1,2,3}$}
\affiliation{%
  $^1$School of Electrical and Computer Engineering, $^2$Department of Physics and Astronomy, and $^3$Purdue Quantum Center \\ Purdue University, West Lafayette, IN  47907
}

\date{\today}% It is always \today, today,

\begin{abstract}
We report our detailed analysis of a table-top system for the measurement of the weak-force-induced electric dipole moment of a ground state hyperfine transition carried out in an atomic beam geometry.  We describe an experimental configuration of conductors for application of orthogonal r.f.\ and static electric fields, with cavity enhancement of the r.f.\ field amplitude, that allows confinement of the r.f.\ field to a region in which the static fields are uniform and well-characterized.  We carry out detailed numerical simulations of the field modes, and analyze the expected magnitude of statistical and systematic limits to the measurement of this transition amplitude in atomic cesium.  The combination of an atomic beam with this configuration leads to strong suppression of magnetic dipole contributions to the atomic signal.  The application of this technique to the measurement of extremely weak transition amplitudes in other atomic systems, especially alkali metals, seems very feasible.  
\end{abstract}

%\pacs{32.70.Cs, 32.80.Qk}% PACS, the Physics and Astronomy Classification Scheme.
\maketitle

\section{Introduction}
Laboratory measurements of very weak atomic transitions that violate the usual parity selection rules are a means of determining the weak force at low collision energies~\cite{BouchiatB75,MarcianoR90,HaxtonW01,Rosner02,DienerGT12}.  The component of this electric dipole transition moment $\mathcal{E}_{\rm PNC}$ that is induced by the weak-force coupling between nucleons has become of great interest in recent years~\cite{FlambaumM97,GomezASOD07,DeMilleCMRK08,SafronovaPJKJS09,TsigutkinDFSYB10,DzubaF12,BorschevskyIDFS13}.  These Nuclear Spin Dependent (NSD) contributions to $\mathcal{E}_{\rm PNC}$ are expected to result from the nuclear anapole moment of the nucleus, with additional smaller contributions from the weak neutral axial-vector nucleon vector electron ($A_n$,$V_e$) current, and the combined effect of the hyperfine interaction and the ($V_n$,$A_e$) current~\cite{JohnsonSS03,SafronovaPJKJS09,FrantsuzovK88,Kraftmakher88,BlundellJS90}.  To date, the only non-zero determination of NSD contributions to $\mathcal{E}_{\rm PNC}$ in \textit{any} element was based upon the difference between measurements of $\mathcal{E}_{PNC}/\beta$ in atomic cesium~\cite{WoodBCMRTW97}, where $\beta$ is the vector polarizability for the transition, on two different hyperfine components of the $6s \rightarrow 7s$ transition; the $F=3 \rightarrow F^{\prime}=4$ and the $F=4 \rightarrow F^{\prime}=3$ lines.  $\mathcal{E}_{PNC}/\beta$ on these lines differed by $\sim$5\% of their average value.  This NSD factor was much larger than was expected, and theoretical efforts~\cite{FlambaumM97,HaxtonW01,JohnsonSS03,AngstmannDF05,SafronovaPJKJS09} to understand this result have not been successful.  Meson exchange coupling constants of the so-called DDH model~\cite{DesplanquesDH80} derived from this result do not agree well with results derived from measurements of the asymmetry in the high-energy scattering of light nuclei~\cite{HaxtonW01,HaxtonLR01,HaxtonLR02,HaxtonH13,Snow15}.  While the applicability of the DDH model to such a large atom is questionable, there is none-the-less strong interest in understanding the NSD of large nuclei, as evidenced by the many efforts underway worldwide in a variety of systems.  Laboratory efforts have sought, or are currently underway, to determine the anapole moment of other nuclei, including Tl~\cite{EdwardsPBN95}, Yb~\cite{Demille95,Kimball01,TsigutkinDFSYB09,TsigutkinDFSYB10}, Fr~\cite{GwinnerGOPSZSBJPAF06,AubinGBPSZCMFSOG11,AubinBCFGGJMOPSSTZZ13}, Ba$^+$~\cite{KoerberSNF03,KleczewskiHSMBF12,WilliamsJHBF13}, Ra$^+$~\cite{VersolatoGWvvJKOSSSTWW10,VersolatoWGvvJKOSSSTWW10,NunezPBBBDGHJMOSSTVWWW13,NunezDMBBGHOSTVWWJ14}, and Yb$^+$~\cite{RahamanDSSZT13}, and several molecular systems as well~\cite{DeMilleCMRK08,BorschevskyIDFS13}.  Differences between $\mathcal{E}_{\rm PNC}$ on various hyperfine lines for these systems could reveal the nuclear anapole moment of these systems.  Comparison between different isotopes of the same species could remove the dependence of the determination on precise atomic theory,  subject to the ability to correct for variations in the nuclear structure among the isotopes~\cite{KozlovPJ01,BrownTR01,DereviankoP02b,GingesF04,BrownDF09}. 

Measurements performed on a hyperfine transition between components of an atomic ground state present an attractive alternative to the above schemes for determining the NSD contributions to $\mathcal{E}_{\rm PNC}$. 
This moment contains only the NSD contribution, simplifying the measurement, and in many cases, the value of $\mathcal{E}_{\rm PNC}$ on ground state transitions is predicted to be larger than the weak amplitude between different electronic states~\cite{DzubaF12}.  Of particular interest is a large program on  francium~\cite{GomezASOD07}, one goal of which is to measure $\mathcal{E}_{\rm PNC}$ on transitions between hyperfine levels of the ground state of this unstable heavy element at TRIUMF.  To carry out these measurements, development of techniques for cooling and trapping these species in a magneto-optical trap (MOT) and carrying out the measurements in this restricted space is necessary.  

The measurement in atomic cesium that we have under development in our laboratory, which we describe in this work, has several features in common with those of the francium effort.  As a ground state transition, atomic coherences are long-lived, and we exploit the interference between the direct transition driven by a radio frequency (r.f.) field and the Raman process driven by a two-frequency cw laser field, in a derivation of the two-pathway coherent control techniques that we have developed for similar measurements~\cite{AntypasE13,AntypasE14}.  Atomic cesium offers several benefits over the francium system that are derived from an atomic beam geometry: that is, a greater atomic density, the capability of sequential preparation, interaction, and detection, and a less restrictive experimental environment.  Furthermore, the beam geometry allow us to spatially separate the interaction regions for the different coherent fields, and to highly suppress the magnetic dipole contributions to the atomic signal, a primary challenge in ground-state measurements of weak signals.  In this work, we discuss how the two-pathway interference method can be used to determine the ratio of the PNC amplitude to the Stark vector polarizability $\beta$.  While our primary interest is in atomic cesium, the technique is generally applicable in any of the stable alkali metal species.

We describe in detail the measurement requirements, and the capability of our technique.  The optimal arrangement uses r.f.\ and static electric fields that are oriented in perpendicular directions, and the r.f.\ field should be confined to a space within which the static field is uniform.  These requirements can be satisfied by a parallel plate transmission line (PPTL) configuration to which cylindrical reflectors (to form an r.f.\ resonant cavity) and isolated conducting pads (for application of the orthogonal d.c.\ field) have been added.  We report the results of our detailed numerical analysis of the electric and magnetic fields supported by this structure, and we use the magnitudes of the field components to estimate the residual systematic effects that one should expect in a determination of $\mathcal{E}_{\rm PNC}$ in atomic cesium.  

This paper is organized as follows.  In Sec.~\ref{sec:CohContSchm}, we discuss the transition probability of a two-level atom interacting with a resonant r.f.\ field and a two-frequency optical field through a Raman interaction.  We show that, when a variable d.c.\ electric field is applied, this coherent control process allows one to determine $\mathcal{E}_{\rm PNC}/ \beta$.  We then discuss in Sec.~\ref{sec:ExpMagnitudes} the various transition amplitudes, including the magnetic dipole, Stark-induced electric dipole, and weak-induced electric dipole, for the transition between hyperfine levels of the ground state of an alkali metal atom.  
We present an estimate of the signal size in Section ~\ref{sec:SignalSize}, with an estimate of the statistical uncertainty, and review the benefits of carrying out the measurement in a standing wave cavity for suppression of magnetic dipole contributions in Section~\ref{sec:StWvCav}.  In the following section, we introduce the PPTL structure, and describe the field modes supported by it.  Finally, we analyze the magnitudes of the dominant residual contributions to the measurement of $\mathcal{E}_{\rm PNC}$, and consider the effects of the distribution of atomic velocities in the beam.

\section{The Coherent Control Scheme} \label{sec:CohContSchm}
We employ the two-pathway coherent control scheme for sensitive measurement of weak moments.  This technique is based on the interference between various optical interactions driven by two or more coherently-related fields.  We developed and employed this technique on measurements of the magnetic dipole transition moment $M$ on the  $6s \: ^2S_{1/2} \rightarrow 7s \: ^2S_{1/2} $ transition in atomic cesium~\cite{AntypasE13,AntypasE14}.  The Fr collaboration bases their measurements on this technique also~\cite{GomezASOD07}.  In this section, we describe the principles behind this technique, with particular attention paid to a transition between hyperfine components of a ground state system, in which both states are long lived.
We show how this measurement can yield a determination of $\mathcal{E}_{\rm PNC}/ \beta$, independent of the profile or amplitude of the r.f.\ field that drives the transition.

\begin{figure}
  \includegraphics[width=5.5cm]{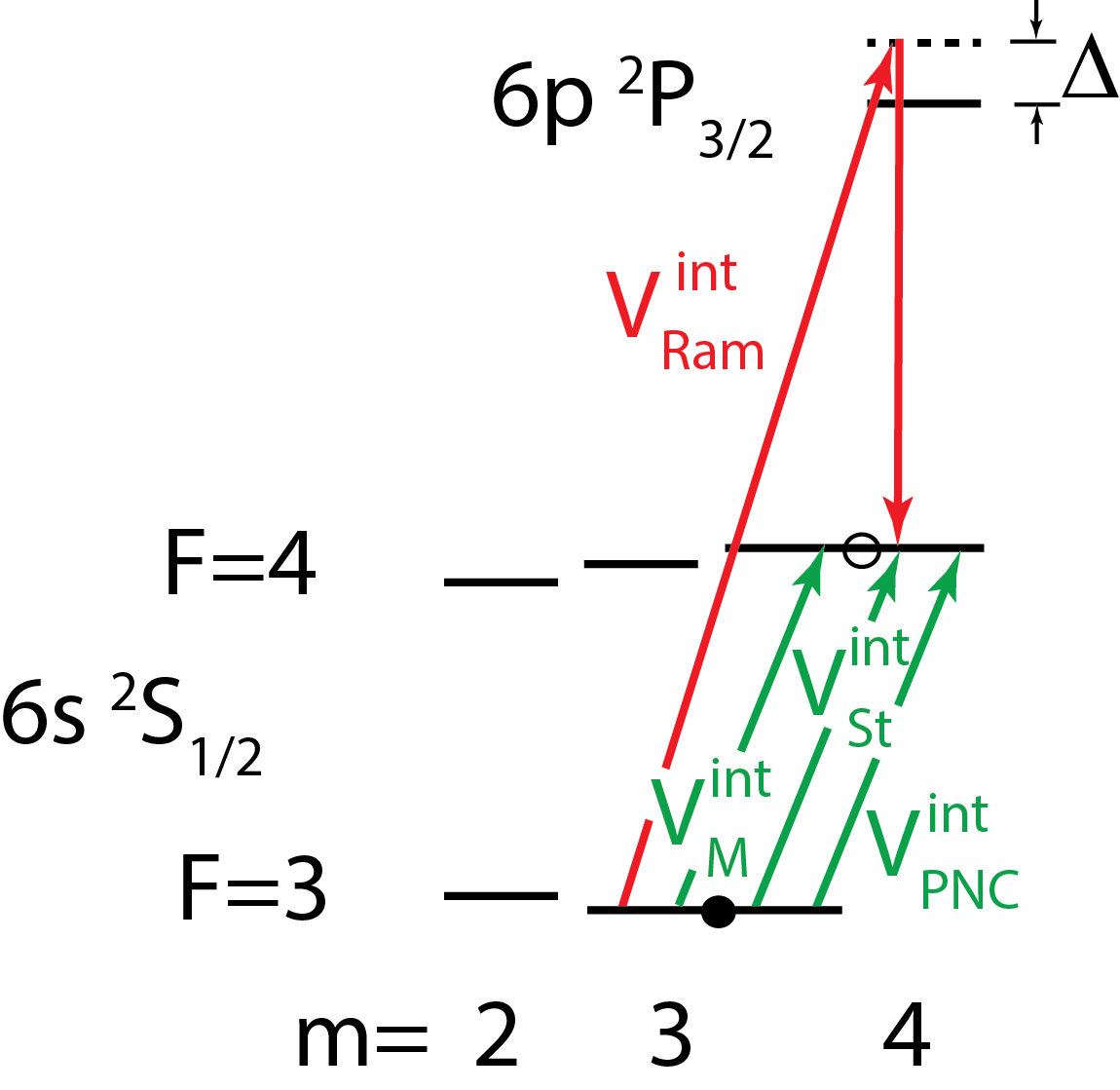}\\
  \caption{(Color online) An abbreviated energy level diagram showing the relevant ground state levels.  We prepare the cesium atoms one hyperfine component of the ground state, $(F,m)$, where $m = \pm F$.  Through the interactions with the r.f.\ field and the optical field, some of the atoms are transferred to the level $(F^{\prime}, m^{\prime})$.  In this figure, we show $(3,3)$ as the initial state, and $(4,4)$ as the final state.   }
  \label{fig:Cs_energy_levels_hfs_ground}
\end{figure}

We consider a sinusoidal wave of amplitude $\varepsilon^{\rm rf}$ and frequency $\omega^{\rm rf} $, incident upon a two-level atom with hyperfine components $ \psi_i $ and $ \psi_f $ of the ground state, of energy $E_i$ and $E_f$, respectively.  We choose the field to be continuous wave, but spatially varying, such that as the atoms move across the interaction region, they effectively see a time-varying field.  When the atoms are initially prepared in a single hyperfine component $\psi_i$, and when the field components are chosen so as to couple the initial state to a single final state $\psi_f$, the atomic system is very closely described as a two-level system, and we can write the state of the atoms using the time-varying amplitudes $c_i(t)$ and $c_f(t)$ as
\begin{displaymath}
   \psi (t) = c_i(t) \psi_i e^{-i \omega_i t} + c_f(t) \psi_f e^{-i \omega_f t}. 
\end{displaymath}
The time evolution of the system is described in terms of the Hamiltonian $H_0 + V^{\rm int}$, where $H_0$ is the atomic Hamiltonian and $V^{\rm int}$ describes the interaction between the atom and the field.  In this work, we consider the weak-force induced electric dipole interaction $V_{\rm PNC}^{\rm int}$, the Stark-induced electric dipole interaction $V_{\rm St}^{\rm int}$, and the magnetic dipole interaction $V_{\rm M}^{\rm int}$ of the atom with the r.f.\ field, plus a Raman interaction $V_{\rm Ram}^{\rm int}$ of the atom with a two-frequency laser field, all of which we describe in more detail later, and write $V^{\rm int}$ as the sum of the individuals
\begin{displaymath}
   V^{\rm int} = V_{\rm PNC}^{\rm int}  + V_{\rm St}^{\rm int} + V_{\rm M}^{\rm int} + V_{\rm Ram}^{\rm int}.
\end{displaymath}
We illustrate these interactions schematically in Fig.~\ref{fig:Cs_energy_levels_hfs_ground}.  

When the atoms exit the interaction region, the probability that they are in state $\psi_f$ is 
\begin{equation}\label{eq:cfeqsinsq}
  |c_f(\infty)|^2 = f(\delta) \sin^2 \left( \left| \sum_i \Theta_i  \right| \right),
\end{equation}
where the $\Theta_i$ are the integrated interaction strengths of any of the individual interactions
\begin{displaymath}
   \Theta_i = \int_{-\infty}^{\infty} \Omega_i(t) \: dt .
\end{displaymath}
The Rabi frequencies of the various interactions are $\Omega_i = V_i^{\rm int}/ \hbar$, and $f(\delta)$ represents the reduction in amplitude when the r.f.\ frequency is detuned from the resonant frequency by $\delta = \omega^{\rm rf} - |E_f - E_i|/ \hbar$.  $f(\delta)$ depends on the temporal shape of the `pulse' as the atoms travel through the interaction region in a non-trivial way, and we will limit our discussion to resonant excitation, $\delta = 0$, for which $f(\delta) = 1$.  

In an atomic beam, collisions are infrequent, and the atoms travel through the interaction region with a constant velocity $v$.  In this case, the interaction strength can be rewritten
\begin{equation}\label{eq:Thetaeq1ovintomegadz}
   \Theta_i = \frac{1}{v}\int_{-\infty}^{\infty} \Omega_i(z) \: dz .
\end{equation}

\begin{figure}[b]
  \includegraphics[width=5cm]{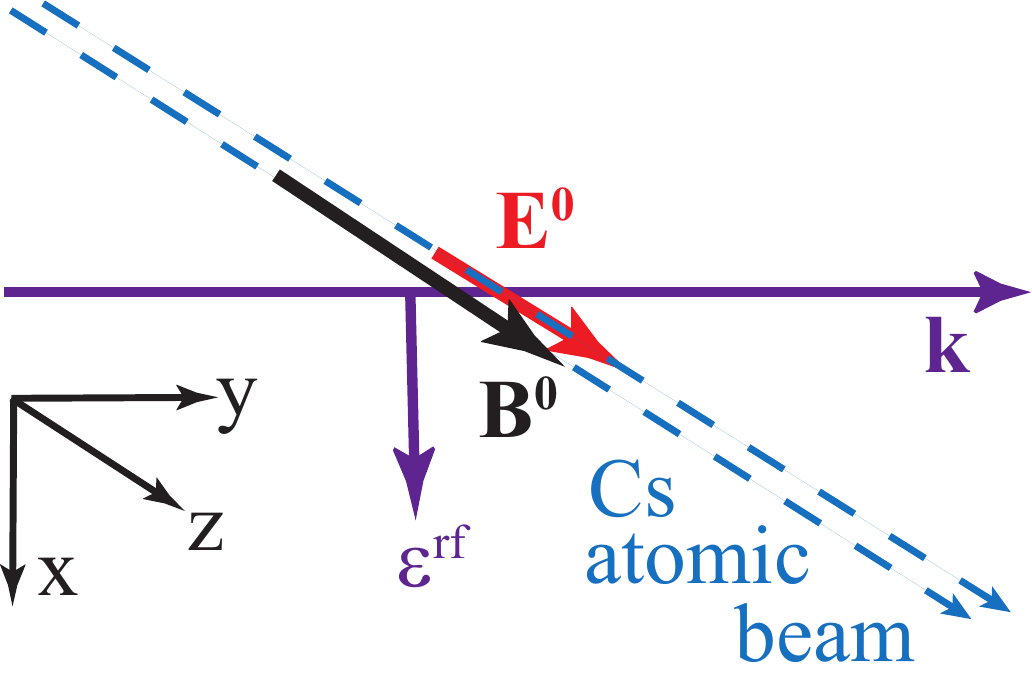}\\
  \caption{(Color online) The field orientations for the measurement of $\mathcal{E}_{\rm PNC}/\beta$ on the $\Delta F = \pm 1$, $\Delta m = \pm 1$ ground state hyperfine transition.  The static electric and magnetic fields are oriented in the $z$-direction, while the polarization of the r.f.\ field is in the $x$-direction.  The polarizations of the laser field components that drive the Raman interaction, not shown, are aligned with the $x$- and $z$-axes.  The r.f.\ and Raman fields propagate parallel to one another, shown as the direction $\mathbf{k}$, as required to maintain a uniform phase difference between interactions.    }
  \label{fig:E_PNC_hfs_ExpGeom_1015}
\end{figure}

We use notation similar to that of Gilbert and Wieman~\cite{GilbertNWW85} for each of the various interactions, and show the optimal field geometry for this measurement in Fig.~\ref{fig:E_PNC_hfs_ExpGeom_1015}.  
That is, the r.f.\ and Raman fields propagate in the $y$-direction, the d.c.\ electric $\mathbf{E}^0$ and magnetic $\mathbf{B}^0$ fields are oriented in the $z$-direction, and the electric field $ \mbox{\boldmath $\varepsilon$} ^{\rm rf}$ of the r.f.\ field is directed in the $x$-direction.  (Parallel propagation of the r.f.\ and Raman fields is necessary to maintain a uniform phase difference between interactions.)  Not shown in this figure are the two components of the laser electric field that drive the Raman transition, each linearly polarized, one in the $x$-direction, the other in the $z$-direction.  In this geometry, the primary r.f.\ and Raman fields each independently drive a $\Delta F = \pm 1$, $\Delta m = \pm 1$ transition, the magnetic dipole contribution on this transition is suppressed, and the Stark-induced and the PNC interactions are in quadrature-phase with one another.  
The primary contributions here, under the precise (idealized) conditions specified in Fig.~\ref{fig:E_PNC_hfs_ExpGeom_1015}, are 
\begin{equation}\label{eq:Astprimary}
  V_{\rm St}^{\rm int} = \beta E_z^0 \varepsilon_x^{\rm rf} \ e^{i ( \omega^{\rm rf}  t - k y - \phi^{\rm rf})}  \ C_{Fm}^{F^{\prime}m \pm 1}
\end{equation}
and 
\begin{equation}\label{eq:APNCprimary}
  V_{\rm PNC}^{\rm int} = \mp i \ Im \{\mathcal{E}_{\rm PNC}\} \varepsilon_x^{\rm rf}  e^{i ( \omega^{\rm rf}  t - k y  - \phi^{\rm rf})}  C_{Fm}^{F^{\prime}m \pm 1}.
\end{equation}
In Eq.~(\ref{eq:Astprimary}), $\beta$ is the vector polarizability and $C_{Fm}^{F^{\prime}m\pm1}$ is a factor related to the Clebsch-Gordon coefficients, defined in detail in Ref.~\cite{GilbertWW84}.  Note that we have explicitly included the phase of the r.f.\ field in these expressions.   

In addition to these primary amplitudes, extra contributions due to magnetic dipole transitions and field misalignments can arise.  The largest of these is 
\begin{eqnarray}\label{eq:AMDelmpm1}
  V_{\rm M}^{\rm int} &= & \eta_0 M \left\{ \rule{0mm}{6mm} \left[ \mp h_x^{\rm rf} + i h_y^{\rm rf} \right]  C_{Fm}^{F^{\prime}m\pm 1}  \right.  \\
  & & \hspace{0.1in} + h_z^{\rm rf}  \left( \frac{\pm B_x^0 + iB_y^0}{B_z^0} \frac{C_{Fm}^{F^{\prime} m} C_{F^{\prime} m \pm 1}^{F^{\prime}m}}{g_{F^{\prime}}} \right. \nonumber  \\
 & &  \left. \left. \hspace{-0.3in} + \frac{\mp B_x^0 + iB_y^0}{B_z^0} \frac{C_{Fm}^{F m\pm 1} C_{F m \pm 1}^{F^{\prime}m \pm 1}}{g_{F}}  \right) \right\} e^{i ( \omega^{\rm rf} t - k y  - \phi^{\rm rf})}  \nonumber 
\end{eqnarray}
for $\Delta m = \pm 1$ transitions, where the $h_i^{\rm rf}$ are the components of the magnetic field of the r.f.\ wave, $M$ is the magnetic dipole transition moment, $\eta_0 = \sqrt{\mu_0 / \epsilon_0} = 120 \pi \: \Omega$ is the impedance of vacuum, and $g_F$ and $g_{F^{\prime}}$ are the gyromagnetic ratio of the initial and final states.  For cesium, $g_F$ is $-1/4$ for the $F=3$ level and $+1/4$ for the $F=4$ level of the ground state.  The first terms in Eq.~(\ref{eq:AMDelmpm1}) are the magnetic dipole amplitude driven by the  $h_x^{\rm rf}$ and $h_y^{\rm rf}$ field components, while the last terms in $h_z^{\rm rf}$ and $B_x^0$ or $B_y^0$ arise from Zeeman mixing of the hyperfine components by the static magnetic field.  To investigate possible interferences from $\Delta m = 0 $ transitions, we also present the magnetic dipole transition amplitude for these transitions
\begin{eqnarray}\label{eq:AMDelm0}
  V_{\rm M}^{\rm int} &= & \eta_0 M \left\{ h_z^{\rm rf}  C_{Fm}^{F^{\prime} m} + \sum_{\pm} \left[ \mp h_x^{\rm rf} + i h_y^{\rm rf} \right] \right.  \nonumber \\
  & & \hspace{-0.3in} \times
    \left[ \left( \frac{\mp B_x^0 + iB_y^0}{B_z^0} \right) \frac{C_{Fm}^{F^{\prime} m \pm1 } C_{F^{\prime} m }^{F^{\prime} m \pm 1 }}{g_{F^{\prime}}}  \right.  \\
 & & \hspace{-0.5in}
  +  \left. \left.  \left( \frac{\pm B_x^0 + iB_y^0}{B_z^0} \right) \frac{C_{Fm}^{F m \mp 1 } C_{F m  \mp 1}^{F^{\prime} m }}{g_{F}}   \right] \right\} e^{i ( \omega^{\rm rf}  t  - k y - \phi^{\rm rf})} .
   \nonumber 
\end{eqnarray}

In addition to these transitions driven by the r.f.\ field, we consider the Raman transition of the form
\begin{displaymath}\label{eq:ARam}
  V_{\rm Ram}^{\rm int} = \tilde{\beta}  \varepsilon_z^{\rm R1} (\varepsilon_x^{\rm R2} )^{\ast}  e^{i ( \omega^{\rm rf}  t - \phi^{\rm Ram} )  }   \ C_{Fm}^{F^{\prime}m \pm 1}
\end{displaymath}
where $\varepsilon_z^{\rm R1}$ and $\varepsilon_x^{\rm R2}$ are the electric field amplitudes of the two laser components, and $\omega^{\rm rf} = \omega^{\rm R1} - \omega^{\rm R2}$, where  $\omega^{\rm R1}$ and $\omega^{\rm R2}$ are the optical frequencies.  The phase $\phi^{\rm Ram}$ is the phase difference between the phases of the two components $\phi^{\rm R1} - \phi^{\rm R2}$.  The Raman polarizability $\tilde{\beta}$ depends on the detuning $\Delta$ of these field components from the D$_2$ transition frequency, and the Raman transition can be enhanced by making $\Delta $ small.

We will analyze these r.f.\ transition amplitudes later using electric and magnetic field amplitudes that we expect to encounter for our parallel plate structure to place limits on unwanted magnetic dipole contributions to the PNC signal.  Before we do this, we return to Eq.~(\ref{eq:cfeqsinsq}), which we examine in the limit of the Raman interaction strength $\Theta_{\rm Ram}$ being much greater than any of the interactions driven by the r.f.\ field $\Theta_{\rm St}$, $\Theta_{\rm M}$, and $\Theta_{\rm PNC}$.  Under these conditions, and with the detuning $\delta = 0$, Eq.~(\ref{eq:cfeqsinsq}) can be expanded to the form
\begin{eqnarray}\label{eq:cfeqdcpmodterm}
   |c_f(\infty)|^2 & = & \sin^2( \left| \Theta_{\rm Ram} \right| ) + \sin(2 \left|  \Theta_{\rm Ram} \right| ) \\
    & & \hspace{-0.5in} \times \sin\left[ \left|  \Theta_{\rm St} + \Theta_{\rm M} + \Theta_{\rm PNC} \right| \cos(\Delta \phi + \delta \phi(E_z)) \right]. \nonumber
\end{eqnarray}
$\Delta \phi = \phi^{\rm rf} - \phi^{\rm Ram}$ is the controllable phase difference between the r.f.\ field and the phase difference $\phi^{\rm Ram}$, and $\delta \phi(E_z) = \tan^{-1}(\mathcal{E}_{\rm PNC} / \beta E_z^0)$ is the phase shift introduced by the quadrature combination of $\mathcal{E}_{\rm PNC}$ and $\beta E_z^0$.  (In writing this phase shift, we presume that the magnetic dipole contributions are suppressed, as we show later.)
We see from this expression a feature that is similar to that of the coherent control scheme on a short-lived state~\cite{AntypasE13,AntypasE14}; that is, that the signal consists of a d.c.\ term resulting from the Raman interaction alone, plus a sinusoidally-varying contribution that varies with the phase difference $\Delta \phi$ between the Raman field and the one-photon r.f.\ field.  Furthermore, the amplitude of the modulating term is the magnitude of the sum of interaction angles $|  \Theta_{\rm St} + \Theta_{\rm M} + \Theta_{\rm PNC} | \approx |  \Theta_{\rm St} + \Theta_{\rm PNC} | $, where we have omitted the small magnetic dipole integrated angle in the final step.  A laboratory measurement of this population modulation amplitude as a function of the d.c.\ electric field $E_z^0$ yields $\mathcal{E}_{\rm PNC}/\beta$.  We see this as follows.  
\begin{displaymath}
  |\Theta_{\rm St} + \Theta_{\rm PNC} | = \frac{1}{ v} \left| \int_{-\infty}^{\infty} \left[ \Omega_{\rm St}(z) + \Omega_{\rm PNC}(z) \right] \: dz \right|,
\end{displaymath} 
which, using Eqs.~(\ref{eq:Astprimary}) and (\ref{eq:APNCprimary}) becomes
\begin{eqnarray}\label{eq:ThetaStplPNC}
  |\Theta_{\rm St} + \Theta_{\rm PNC} | & = & \frac{1}{\hbar v} \left| \beta E_z^0 \mp i  \ Im \{\mathcal{E}_{\rm PNC}\} \right| \nonumber \\  
   & & \hspace{-0.3in} \times C_{Fm}^{F^{\prime}m \pm 1} \int_{-\infty}^{\infty} \varepsilon_x^{\rm rf} (z) \: dz,
\end{eqnarray} 
valid when $E_z^0$ is uniform in the interaction region.
Since the Stark and PNC moments add in quadrature, the amplitude of the sinusoidal modulation of the signal scales as 
\begin{equation}\label{eq:QuadSumtplPNC}
   \left| \beta E_z^0 \mp i  \ Im \{\mathcal{E}_{\rm PNC}\} \right| = \sqrt{ (\beta E_z^0)^2 + \left| \mathcal{E}_{\rm PNC} \right|^2}.
\end{equation} 
At small d.c.\ field, the modulation amplitude is proportional to $Im \{\mathcal{E}_{\rm PNC}\}$ alone, while at large field, the modulation amplitude is nearly proportional to $\beta E_z^0$.  By measuring this amplitude of the population modulation as a function of the d.c.\ field, therefore, one can determine the ratio $\mathcal{E}_{\rm PNC} /\beta$.  

To optimize the amplitude of the signal modulation in Eq.~(\ref{eq:cfeqdcpmodterm}), one should adjust the amplitude of the Raman interaction to $| \Theta_{\rm Ram} |= \pi/4$.  At this value, the factor $\sin(2 \left|  \Theta_{\rm Ram} \right| )$ is equal to 1, and the atomic population due to the Raman interaction alone is equal to 1/2, i.e. equal probability in the initial and final states.  Any additional interactions of the atom with the r.f.\ field add (slightly) to the population in the $\psi_f$ state when this interaction is in phase with the Raman interaction, and subtract when out-of-phase.

\begin{figure}
  \includegraphics[width=7.8cm]{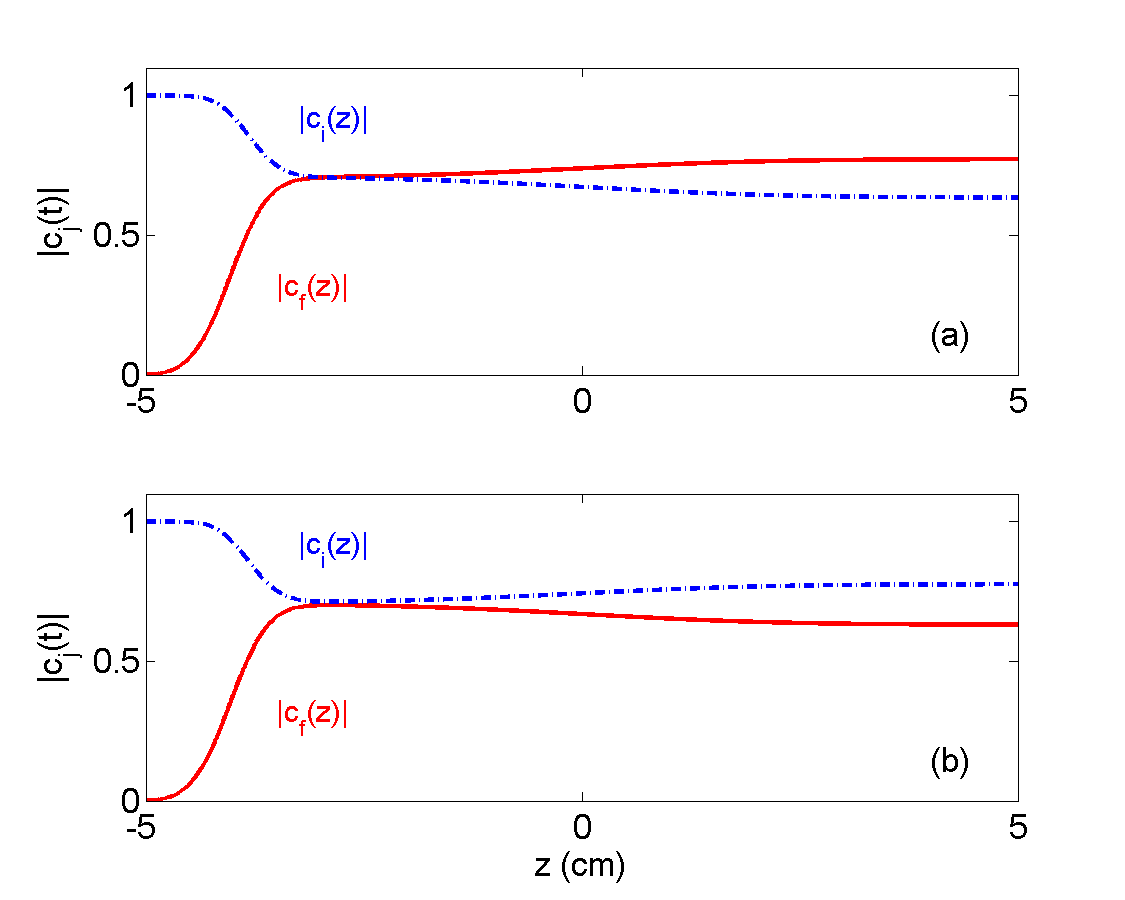}\\
  \caption{(Color online) The variation of state amplitudes $|c_f(z)|$ (red solid) and $|c_i(z)|$ (blue dashed) versus $z$ as the atoms pass through the interaction region from left to right.  The atoms are prepared by the Raman beams in a superposition state before entering the broad r.f.\ field.    Both fields are Gaussian in shape, with peak Rabi frequency and beam radii of $\Omega_{\rm Ram,0}$ = 23.9 ms$^{-1}$ and 0.5 cm for the Raman beam, and $\Omega_{\rm w,0}$ = 0.61 ms$^{-1}$ and 2.5 cm for the r.f.-driven interaction. 
  In (a), the Raman and r.f.\ interactions are in phase with one another, while in (b), the interactions are out of phase.  In either case, the duration of the interaction is $w_{\rm Ram}/v \simeq$ 19 $\mu$s for the Raman beam, and $w_{\rm rf}/v \simeq$ 93 $\mu$s for the r.f.\ field.  }
  \label{fig:Cevstime}
\end{figure}

\begin{figure}
  \includegraphics[width=7.8cm]{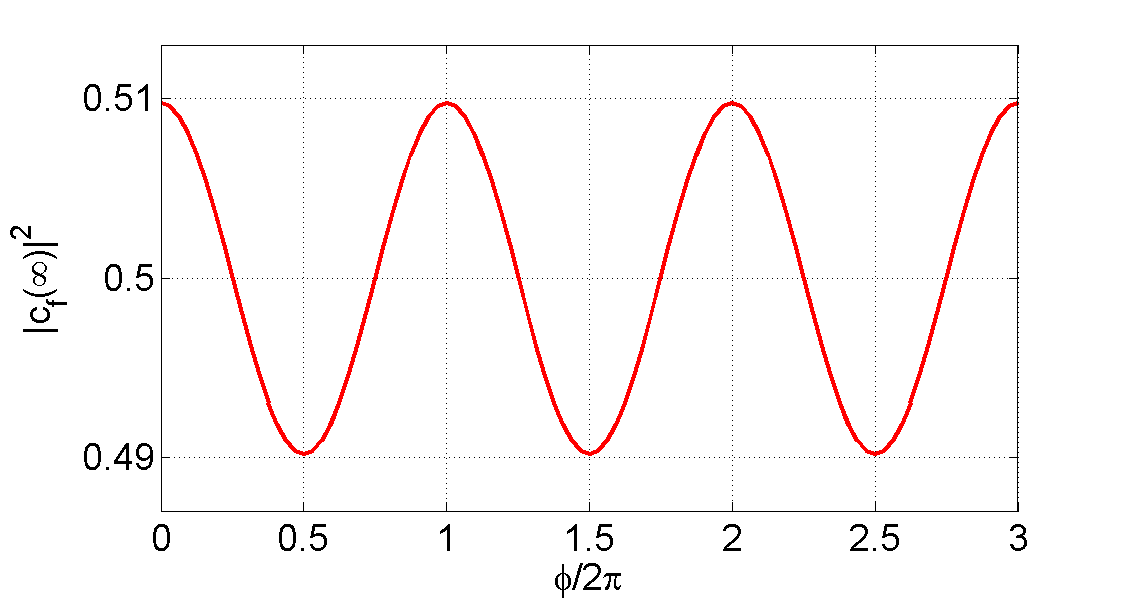}\\
  \caption{(Color online) The sinusoidal variation of the signal as a function of the phase difference between the r.f.\ and Raman interactions.  The peak Rabi frequency of the r.f.-driven interaction is $\Omega_{\rm w,0}$ = 0.061 ms$^{-1}$ for this plot.  Other parameters are as given in the caption to Fig.~\protect\ref{fig:Cevstime}.    }
  \label{fig:SigvsPhase}
\end{figure}

We can gain some insight into the interference by following the evolution of the amplitudes $|c_f(t)|$ (red solid) and $|c_i(t)|$ (blue dashed) as the atoms move across the interaction region, which we show in Fig.~\ref{fig:Cevstime}.  For this illustration, the atoms move from left to right, and encounter the Raman field first, centered at $z$ = -4 cm, which prepares them in a coherent superposition state.  The atoms then enter the broad r.f.\ field.  We use Gaussian profiles for the r.f.\ and Raman fields.  For the former, the peak amplitude is $\varepsilon_{x,0}^{\rm rf}$ and beam radius $w_{\rm rf}$ in the interaction region, 
\begin{displaymath}
   \varepsilon_x^{\rm rf}(z) = \varepsilon_{x,0}^{\rm rf} \  e^{- (z / w_{\rm rf})^2}.
\end{displaymath}
We show this for two values of the phase $\Delta \phi$ in 
Fig.~\ref{fig:Cevstime}.
Fig.~\ref{fig:Cevstime}(a) shows the magnitudes of the state amplitudes when the Raman and r.f.-driven interactions are in phase with one another, while Fig.~\ref{fig:Cevstime}(b) shows the state amplitudes when the interactions are $\pi$ out of phase with one another.  The peak Rabi frequency, center position, and beam radius are $\Omega_{\rm Ram,0}$ = 23.9 ms$^{-1}$, $z_c$ = -4 cm, and $w_{\rm Ram}$ = 0.5 cm for the Raman beam, and $\Omega_{\rm w,0}$ = 0.61 ms$^{-1}$, $z_c$ = 0, and $w_{\rm rf}$ = 2.5 cm for the r.f.-driven interaction.  We use 270 m/s, the peak velocity of the atoms  in our atomic beam for $v$.  The duration of the interaction is $w_{\rm Ram}/v \simeq$ 19 $\mu$s for the Raman beam, and $w_{\rm rf}/v \simeq$ 93 $\mu$s for the r.f.\ field.  When the amplitudes are in phase with one another, $|c_f(z)|$ grows monotonically, while when the interactions are out of phase, the amplitude decreases after its initial preparation by the Raman beam.  The value of $|c_f(\infty)|$ after the atoms have exited the interaction region is $\sqrt{1/2 + \sin ( | \Theta_w | )}$ for in-phase interactions and $\sqrt{1/2 - \sin ( | \Theta_w | )}$ for out-of-phase interactions.  
When the PNC and Stark-induced terms are driven by the r.f.\ field, then $| \Theta_w |$ is $|\Theta_{\rm St} + \Theta_{\rm PNC} |$, where the PNC interaction angle is 
\begin{eqnarray}\label{eq:Theta_PNC}
  \Theta_{\rm PNC} & = & \left( \mp i Im \{\mathcal{E}_{\rm PNC}\}    C_{Fm}^{F^{\prime}m \pm 1} / \hbar v \right) \int_{-\infty}^{\infty} \varepsilon_x^{\rm rf} (z) \: dz \nonumber \\
     & & \hspace{-0.5in} =  \left( \mp i Im \{\mathcal{E}_{\rm PNC}\}    C_{Fm}^{F^{\prime}m \pm 1} / \hbar v \right) \sqrt{\pi} \ w_{\rm rf}   \ \varepsilon_{x,0}^{\rm rf}. 
\end{eqnarray}
Similarly, the integrated area of the Stark-induced interaction angle for this Gaussian-shaped profile is $\Theta_{\rm St} = \beta E_z^0 C_{Fm}^{F^{\prime}m \pm 1} \sqrt{\pi} \ w_{\rm rf}   \ \varepsilon_{x,0}^{\rm rf} / \hbar v $.  
The term $1/2$ in the expressions for $|c_f(\infty)|$ comes from $\sin^2(| \Theta_{\rm Ram} | )$ with $ | \Theta_{\rm Ram} |  = \sqrt{\pi} w_{\rm Ram}  | \Omega_{\rm Ram, 0} |  / v$.  The weak signal strength is $ | \Theta_w | = \sqrt{\pi} w_{\rm rf}  | \Omega_{\rm w, 0} |  / v $ in this example is 0.10.   
Any interaction of the atoms with the r.f.\ field therefore is evident as a modulation of this signal as we vary the phase difference between the fields.  We illustrate this in Fig.~\ref{fig:SigvsPhase}, which shows the sinusoidal modulation of the final state population as a function of $\Delta \phi$.  
Here the parameters are as they were in Fig.~\ref{fig:Cevstime}, with the exception of $\Omega_{\rm w,0}$ which we have decreased to 0.061 ms$^{-1}$ for this figure. The amplitude of the modulation of $|c_f(\infty)|^2$ is $| \Theta_w | = \sqrt{\pi} w_{\rm rf} \Omega_{\rm w,0} /v = 0.010$, in agreement with the numerical data in the figure.  In our simulations, the amplitude of the modulation scales linearly with the weak amplitude.

Important conditions and features of this measurement technique include:
\begin{enumerate}
  \item Mutual coherence of the different time-varying fields is required.  This can be implemented in the laboratory by using non-linear mixing, injection locking of diode lasers, or frequency modulation techniques.
  
  \item The coherent beams that drive the interactions must propagate in the same direction in order to maintain a uniform phase difference for all atoms in the interaction region.   
  
  \item The Raman and the r.f.\ field distribution need not overlap one another. 
  Since the ground state is long lived, the final level retains its coherence, and the net excitation of the final state depends on the accumulated effect across the interaction region.  
    
  \item We control the phase difference between the transition amplitudes with r.f.\ devices, completely external to the interaction region.
  
  \item We select the particular interactions that contribute to the measurement by choosing the orientation of the various fields in the interaction region.
  
    \item The measurement uses only modest d.c.\ electric fields, $\lesssim$ 100 V/cm.  This allows flexibility in the experimental configuration.  
  
  \item Since the interactions $\Omega_{\rm PNC}$ and $\Omega_{\rm St}$ are $\pi/2$ out of phase with one another, these amplitudes add in quadrature.  This indicates that the amplitude of the modulating signal is at a minimum when the static electric field is turned off, and increases when a static field of either polarity is applied.
  
  \item Using different field orientations, this coherent control technique may be used to determine $M/\beta$.  This may be a useful means of determining the vector polarizability $\beta$, but we defer any further discussion of this to a future report. 
\end{enumerate}

In the following sections, we will discuss the expected magnitudes of the different interactions, and present an experimental assembly of conductors for such a measurement in an atomic beam configuration.  Finally, we will analyze the effect of expected magnetic dipole contributions to the measurement.

\section{Expected magnitudes of $M$, $\beta$, and $\mathcal{E}_{\rm PNC}$}\label{sec:ExpMagnitudes}

In order to design a measurement system and understand the effect of stray fields and the magnitude of unwanted contributions to the signal, we must first know the expected magnitudes of the PNC moment, $\mathcal{E}_{\rm PNC}$, the vector polarizability $\beta$, and the magnetic dipole moment $M$ for the transition.  

The PNC amplitude for this transition is calculated~\cite{DzubaF12} to be 
\begin{equation}\label{eq:EPNC}
   \mathcal{E}_{\rm PNC} = 1.82 \times 10^{-11} i e a_0, 
\end{equation}
where $e$ and $a_0$ are the electron charge and the Bohr radius, respectively.  This is larger than $\mathcal{E}_{\rm PNC}$ for the moment on the $6s \rightarrow 7s$ transition in cesium by a factor of 2.2.

The vector polarizability has not previously been calculated, but we can estimate its approximate magnitude using the sum-over-states expansion of Refs.~\cite{BouchiatB75} and \cite{GilbertWW84},
\begin{eqnarray*}
  \beta & = & \frac{e}{6 \hbar } \left[ \sum_n  r_{n, 1/2}^2   \left(\frac{1}{\Delta_{4; n,1/2}}  - \frac{1}{\Delta_{3; n,1/2}}  \right) \right.  \\
    & & \left. \hspace{0.4in}  + \frac{1}{2} r_{n, 3/2}^2   \left(\frac{1}{\Delta_{4; n,3/2}}  - \frac{1}{\Delta_{3; n,3/2}} \right) \right],
\end{eqnarray*}
where $r_{n, j}$ represents the reduced dipole matrix elements $\langle np_{j}|| r || 6s_{1/2} \rangle$ for $j = 1/2$ or $3/2$, and $\hbar \Delta_{F;n,j}$ are the energy differences $E_{6s,F} - E_{np_j}$ for the two hyperfine states $F=3$ or $4$ of the ground $6s \: ^2S_{1/2} $ and the excited $np \: ^2P_j $ states.  The $n=6$ term dominates this sum, and the ground state hyperfine splitting $\Delta_{\rm hfs}$ is small compared to the energy of the 6p states, so the polarizability is approximately  
\begin{eqnarray*}
   \beta & \simeq & \frac{e \Delta_{\rm hfs}}{6} \left[ \frac{\left| \langle 6p_{1/2}|| r || 6s_{1/2} \rangle \right|^2}{(E_{6s} - E_{6p_{1/2}})^2} \right. \\ 
  & & \hspace{0.7in} \left. + \frac{1}{2} \frac{\left| \langle 6p_{3/2}|| r || 6s_{1/2} \rangle \right|^2}{(E_{6s} - E_{6p_{3/2}})^2} \right]  . \nonumber
\end{eqnarray*}
We use $\langle 6p_{1/2}|| r || 6s_{1/2} \rangle = 4.5062 \: a_0$ and $\langle 6p_{3/2}|| r || 6s_{1/2} \rangle = 6.3400 \: a_0$~\cite{YoungHSPTWL94,RafacT98,RafacTLB99,AminiG03,DereviankoP02,BouloufaCD07} to estimate the vector polarizability for this transition as $\beta \simeq 0.00346 \: a_0^3$.  Based on these expected magnitudes of $\beta$ and $\mathcal{E}_{\rm PNC}$, the ratio $\mathcal{E}_{\rm PNC}/\beta$ is about 27 V/cm; upon application of a static electric field of this magnitude, the magnitudes of the Stark-induced amplitude and the PNC amplitude are equivalent. Since $\beta$ is so small for this transition, we conclude that systematic errors due to uncontrolled electric fields in the interaction region, due to surface contamination and patch effects and estimated to be $\lesssim$0.1 V/cm, are inconsequential in these ground state measurements.  This is in strong contrast to measurements of $\mathcal{E}_{\rm PNC}$ on the $6s \rightarrow 7s$ transition~\cite{WoodBCMRTW97}, for which uncontrolled electric fields were of major concern.

In addition to these two relatively weak amplitudes driven by the r.f.\ field, the magnetic dipole moment is active on this transition.  The amplitude for this transition is $V_{\rm M}^{\rm int} = \langle 6s \: ^2S_{1/2} \: F^{\prime} m^{\prime} | - \mbox{\boldmath $\mu$}_m  \cdot \mathbf{b}^{\rm rf} |  6s \: ^2S_{1/2} \:  F m \rangle$, where $\mbox{\boldmath $\mu$}_m = \mu_B (g_L \mathbf{L} + g_S \mathbf{S} + g_I \mathbf{I})$ is the magnetic moment of the atom, $\mu_B = e \hbar /2m$ is the Bohr magneton, and $\mathbf{b}^{\rm rf}$ is the magnetic flux density of the r.f.\ wave.  $\mathbf{L}$, $\mathbf{S}$ and $\mathbf{I}$ are the usual orbital, spin, and nuclear angular momenta, and $g_L$, $g_S$, and $g_I$ are the respective gyromagnetic ratios.  For the transition of this work, the orbital angular momentum is zero, and $g_I$ is much less than $g_S$ (which is $\approx 2$) due to the heavy mass of the nucleus.  For the ground state transition, the spatial parts of $\psi_i$ and $\psi_f$ are the same, and using $\varepsilon^{\rm rf} / b^{\rm rf} = c$, the magnetic dipole amplitude is $M = \mu_B g_S / 2 c \simeq \mu_B/c $.  But $\mu_B/c = e a_0 \alpha /2$, where $\alpha \simeq 1/137$ is the fine structure constant, so $M \simeq e a_0 \alpha /2$, and the ratio $M / \mathcal{E}_{\rm PNC} \simeq 2 \times 10^8$.    
The magnetic dipole contributions to the signal must be suppressed for a successful measurement of $\mathcal{E}_{\rm PNC}$, representing the primary challenge of these measurements.  The orientations of the field components that we have shown in Fig.~\ref{fig:E_PNC_hfs_ExpGeom_1015} are an important first step in meeting this challenge.

\section{Magnitude of Signal} \label{sec:SignalSize}
In this section, we will use the results of the analysis of Sec.~\ref{sec:CohContSchm}, in particular Eqs.~(\ref{eq:cfeqdcpmodterm}) and (\ref{eq:Theta_PNC}), and the calculated value of $\mathcal{E}_{\rm PNC}$ given in Eq.~(\ref{eq:EPNC}), to estimate the magnitude of the PNC signal, and from this the integration time required to achieve a useful statistical uncertainty of the measurement.  To calculate the signal size, we will use $|C_{Fm}^{F^{\prime}m \pm 1}| =  \sqrt{7/8}$, $\varepsilon_{x,0}^{\rm rf} $ = 250 V/cm, and $w_{\rm rf}$ = 2.50 cm.  The value of $C_{Fm}^{F^{\prime}m \pm 1}$ is valid for cesium ground state transitions $(F, m) = (3, \pm 3) \rightarrow (4, \pm 4)$ or $(4, \pm 4) \rightarrow (3, \pm 3)$, and we will show in Sec.~\ref{sec:PPTL} that the values of the peak field amplitude and radius are reasonable.  Then using the cesium atomic beam peak velocity $v$ = 270 m/s, we estimate that the interaction angle for the PNC interaction is
\begin{displaymath}\label{eq:ThetaPNCestmag}
   \Theta_{\rm PNC} = \pm i 5.6 \times 10^{-6}.
\end{displaymath}
To measure this amplitude, one can drive the interfering Raman and PNC interactions, and count the transition rate as a function of the phase difference between the transitions.  A minimal measurement may consist of $N_+$, the total count of atomic excitations when the r.f.\ and Raman interactions are in phase with one another ($|c_f|^2 = \frac{1}{2} + |\Theta_{\rm PNC}|$), and $N_-$ the total count of excitations when the r.f.\ and Raman interactions are $\pi$ out of phase with one another ($|c_f|^2 = \frac{1}{2} - |\Theta_{\rm PNC}|$).  Then
\begin{displaymath}
    \Theta_{\rm PNC} = \frac{1}{2} \ \frac{N_+ - N_-}{N_+ + N_-}.
\end{displaymath} 
To use this result to determine $\mathcal{E}_{\rm PNC}$, however, one must also have an accurate determination of the r.f.\ beam profile and field amplitude.  Alternatively, one can apply a d.c.\ electric field to the atoms, and measure the amplitude of the modulation as a function of the field amplitude $E_z^{\rm 0}$, as suggested in Eqs.~(\ref{eq:ThetaStplPNC}) and (\ref{eq:QuadSumtplPNC}).  

When the precision of $N_+$ and $N_-$ is limited by counting statistics, then the uncertainty in either of these counts is $\sigma_N = \sqrt{N}$, where $N$ represents either $N_+$ or $N_-$ (which are essentially the same).  The uncertainty in $\Theta_{\rm PNC}$ is $\sigma_{\rm PNC} = 1/ \sqrt{8N}$, and to achieve a 3\% measurement of $\Theta_{\rm PNC}$, one must count $N = 1/8\sigma_{\rm PNC}^2 = 3 \times 10^{12}$ atoms for each individual measurement.  In a counting interval $T$, the number of counts is $N = \frac{1}{2} \rho_{\rm Cs} A v T$, where $\frac{1}{2}$ is the average excitation probability, $\rho_{\rm Cs}$ is the number density of the atomic beam ($10^9$ cm$^{-3}$), $A$ is the cross sectional area of the atomic beam (1 mm$^2$), and $v$ is the peak velocity of the atoms in the beam.  The counting time T to achieve the required statistics is 20 seconds per data point.  During the course of a measurement, one must repeat the process at many different phases, not just two, and one must vary the d.c.\ electric field strength $E_z^0$ and repeat the measurement.  Regardless, the estimate of the integration time shows that the measurement is feasible in the beam geometry.  

We conclude this section with an estimate of the maximum value of the d.c.\ field amplitude $E^0$ needed.  As discussed in the previous section, we expect that the ratio $\mathcal{E}_{\rm PNC} / \beta$ is approximately 27 V/cm.  In carrying out the measurements, we must vary the Stark-induced angle $\Theta_{\rm St}$ over the range from zero to $\sim \pm 3 |\Theta_{\rm PNC}|$.  This requires a variable field strength of maximum value $3\mathcal{E}_{\rm PNC} / \beta \approx \pm $80 V/cm.   

\section{Standing Wave Cavity} \label{sec:StWvCav}
In the previous section, we estimated the magnitude of the hyperfine ground state PNC coherent control signal, based on expected atomic parameters and reasonable field strengths that can be generated in the laboratory.  Among the latter was an r.f.\ field amplitude $\varepsilon_{x,0}^{\rm rf} $ of 250 V/cm.  This field amplitude can be achieved either inside a resonant power build-up cavity, or by using a very large r.f.\ amplifier.  Use of a resonant cavity also helps to suppress the magnetic dipole contributions to the measured signal, as we now discuss.  This approach is also discussed in Ref.~\cite{GomezASOD07}.  

\begin{figure}
  \includegraphics[width=5.5cm]{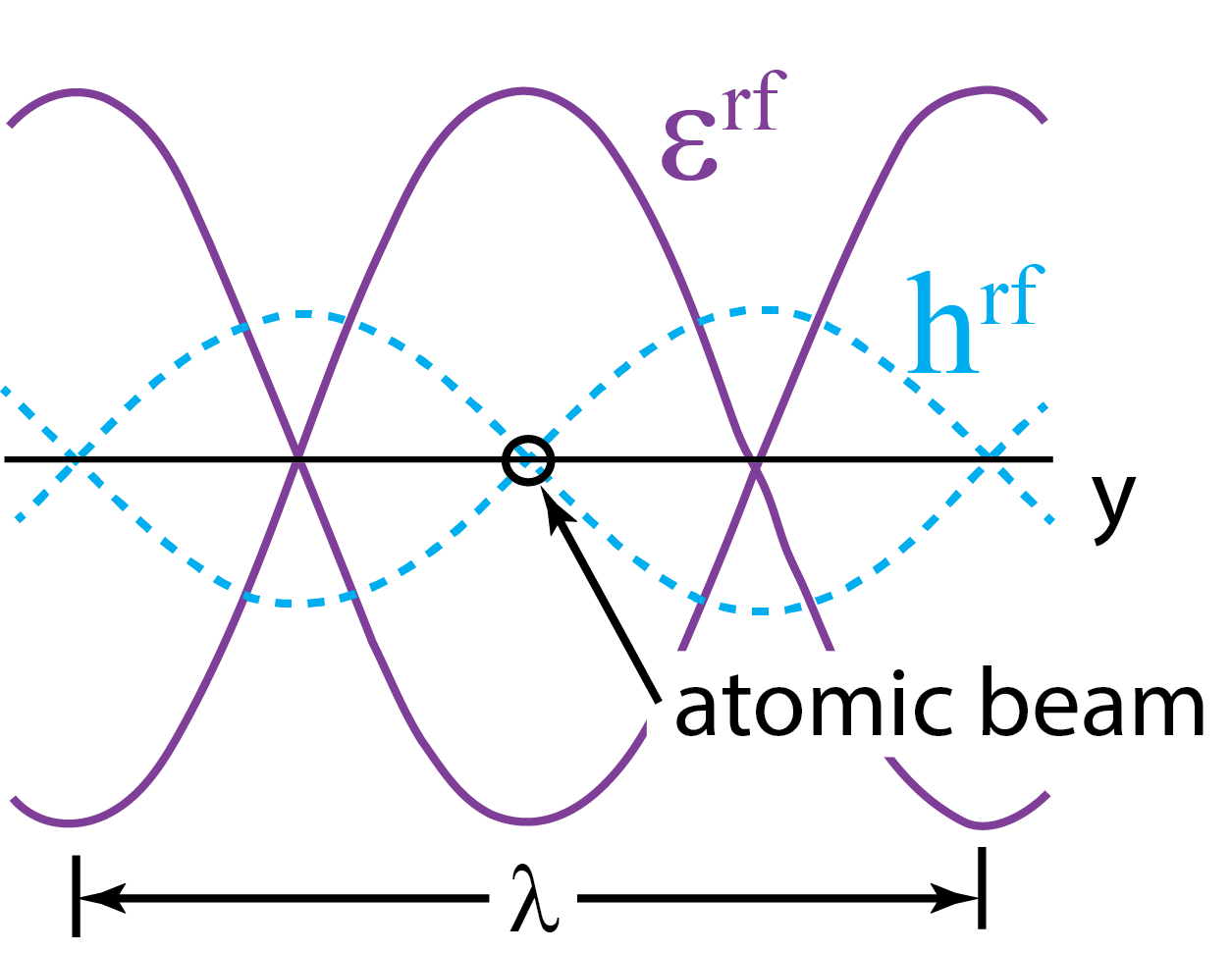}\\
  \caption{(Color online) The standing wave pattern of the r.f.\ electric field $\varepsilon_x^{\rm rf}$ and magnetic field $h_z^{\rm rf}$, with the atomic beam located at the node of the magnetic field.   }
  \label{fig:BEeps_orientations_SWP}
\end{figure}

As we discussed earlier, the large magnetic dipole amplitude is suppressed to first order by the choice of orientations of the primary fields.  (The $h_z^{\rm rf}$ component drives a $\Delta m = 0$ transition, whereas the interference that we have discussed takes place on a $\Delta m = \pm 1 $ transition.)  Still, due to the large magnitude of the ratio $M / \mathcal{E}_{\rm PNC} $ and reasonable limits in the field uniformity and experimental alignment, additional measures are required to suppress this interaction further.  This additional suppression can be achieved by working in a standing wave configuration, in which the nodes of the magnetic field coincide with the anti-nodes of the electric field, as we illustrate in Fig.~\ref{fig:BEeps_orientations_SWP}.
At this point, the interactions $V_{\rm PNC}^{\rm int}$ and $V_{\rm St}^{\rm int}$ are maximized, and $V_{\rm M}^{\rm int}$ is minimized.  To take best advantage of this, one should (1) use a cavity geometry in which the amplitudes of the traveling waves propagating in the $+y$ and $-y$ directions, $\varepsilon_+^{\rm rf}$ and $\varepsilon_-^{\rm rf}$, respectively, are equal, and (2) keep the radius $b$ of the atomic beam small.  The first requires either that the cavity is symmetric (the reflectivities of the two end reflectors are equal, and the cavity is excited by equal amplitude inputs on each side), or that one of the reflectors has unit reflectivity.  The choice of beam radius $b$ is a compromise between large atom number, improving the counting statistics, or small magnetic dipole amplitude for atoms at the edge of the beam, which scales as $\sin(kb) = \sin(2 \pi b/\lambda)$, where $\lambda$ = 3.2 cm is the wavelength of the 9.2 GHz wave.  For $b$ = 0.5 mm, this reduction factor is $\sim$0.1.  Furthermore, the sign of the magnetic dipole amplitude is opposite on the two sides of the node, further reducing this contribution.  We will return to this reduction in Section~\ref{sec:systerrors}. 
In the next section, we will discuss the design and analysis of a symmetric r.f.\ power build-up cavity based on a parallel plate transmission line structure, which allows spatial confinement of the r.f.\ field and generation of a transverse d.c.\ electric field.

\section{Parallel Plate Transmission Line Structure}\label{sec:PPTL}
The measurement that we have described  presents several experimental challenges.  First, we must apply r.f.\ and static electric fields that are oriented in directions that are perpendicular to one another.  Second, we require that the r.f.\ field is in a standing wave configuration for suppression of the magnetic dipole contributions.  And third, we must minimize the unwanted field components of the r.f.\ field, as these also lead to systematic magnetic dipole contributions to the signal. In this section, we describe an electrode configuration that allows us to meet these requirements.  

\begin{figure}
  \includegraphics[width=7.9cm]{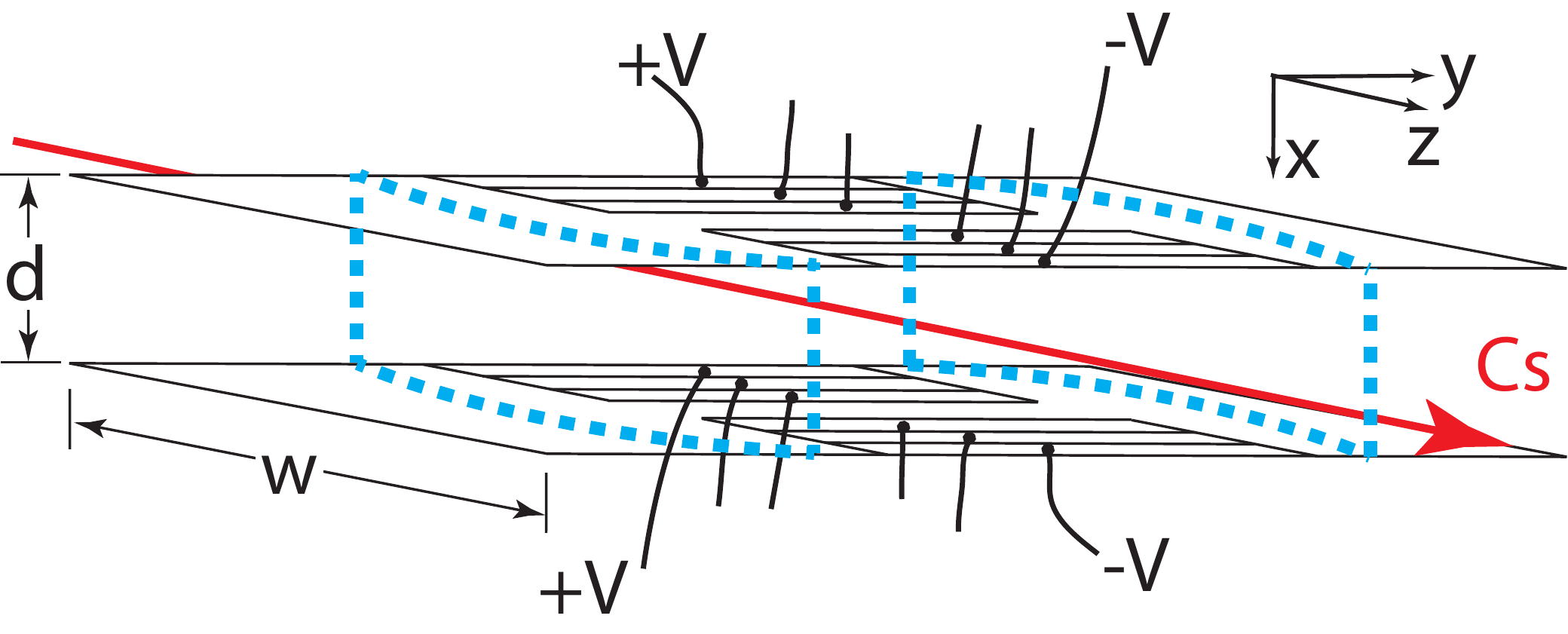}\\
  \caption{(Color online) The electrode configuration that supports the standing wave r.f.\ field $\varepsilon_x$ and the static electric field $E_z^0$.   }
  \label{fig:PPTL_structure_w_ref_rotated}
\end{figure}
In Fig.~\ref{fig:PPTL_structure_w_ref_rotated},  
we show a section of a parallel plate transmission line, with waves propagating in the $\pm y$ directions, that is modified in two regards.  First, we have isolated several conducting pads on the top and bottom conductors for application of a d.c.\ bias, and secondly, we have inserted cylindrical reflectors to either side of the interaction region to form an r.f.\ cavity, open on the $z$ faces, allowing power build-up of the cavity mode at the resonant frequency.  When we have biased the d.c.\ pads progressively, at a voltage $+V$ on one side to $-V$ on the other, we can generate an electric field $E^0$ in the central region between the plates that is primarily directed in the $\pm z$-direction. We capacitively couple each of the bias pads to the transmission line structure so that they carry the a.c.\ components without any significant perturbation.  For a transmission line characteristic impedance $Z_0$ = 50 $\Omega$, this requires a coupling capacitance of $C \gtrsim$ 30 pF.

We can model the cavity modes that are supported by the parallel-plate structure in the region between the cylindrical reflectors approximately using the elliptical Hermite-Gaussian modes as described in Yariv~\cite{Yariv85}.  These modes are nearly Gaussian in shape in the $z$-direction, but uniform in the $x$-dimension, in the limit of an infinite beam size in this dimension. Within the cavity, the spatial mode is described by the superposition of waves traveling in the $+y$ and $-y$ directions, 
\begin{equation}\label{eq:exSWfoc}
  \varepsilon_x^{\rm rf}(y,z) = \varepsilon_{+}^{\rm rf}(y,z)  + \varepsilon_{-}^{\rm rf}(y,z),
\end{equation}
and 
\begin{equation}\label{eq:hzSWfoc}
  h_z^{\rm rf}(y,z) = \frac{1}{\eta_0} \left( \varepsilon_{+}^{\rm rf}(y,z)  - \varepsilon_{-}^{\rm rf}(y,z) \right),
\end{equation}
where
\begin{eqnarray*}
  \varepsilon_{\pm}^{\rm rf}(y,z) & = & \varepsilon_{0,\pm}^{\rm rf} \sqrt{\frac{w_0}{w(y)}} \exp \left\{ \rule{0cm}{0.5cm} \mp i \left[ ky - \eta(y) \right] \hspace{0.2in} \right. \\
  & & \left. \hspace{0.9in} -z^2 \left[ \frac{1}{w^2(y)} + \frac{ik}{2R(y)} \right] \right\} , \nonumber
\end{eqnarray*}
In these expressions, $w_0$ is the $1/e^2$ (intensity) beam radius at the focus, the beam profile radius a distance $y$ from the focus is 
\begin{displaymath}
  w(y) = w_0 \: \sqrt{ 1 + (y/y_0)^2 }  ,
\end{displaymath}
$y_0$ is the confocal parameter 
\begin{displaymath}
   y_0 = \pi w_0^2/\lambda,
\end{displaymath}
$R(y)$ is the radius of curvature of the wavefronts 
\begin{displaymath}
   R(y) = y \left[ 1 +  (y_0/y)^2 \right], 
\end{displaymath}
and $\eta(y)$ 
\begin{displaymath}
   \eta(y) = \frac{1}{2} \tan^{-1} \left( y/y_0 \right) 
\end{displaymath}
is the slow phase shift (the Guoy phase) through the focal region.  For a symmetric cavity constructed of cylindrical reflectors of radius of curvature $R$ separated by a distance $\ell_c$, the confocal parameter is $y_0 = (\ell_c/2)\sqrt{2R/\ell_c - 1}$, the beam radius at the center is $w_0 = (\lambda \ell_c/2\pi)^{1/2} (2R/\ell_c - 1)^{1/4}$, and the beam radius at the reflectors is $w(y = \pm \ell_c /2) = (\lambda R/\pi)^{1/2} (2R/\ell_c - 1)^{-1/4}$.  
The cavity mode has an electric field anti-node (and magnetic field node) at the center when the cavity length $\ell_c$ is approximately $(n+1/2) \lambda$, where $n$ is an integer.  The r.f.\ beam radius $w(y = \pm \ell_c/2)$ at the reflectors is minimized when the reflector spacing is confocal, i.e. $\ell_c = R$.  By adjusting the reflector slightly away from the confocal spacing, one can retain the small mode size $w(\pm \ell_c/2)$ at the reflectors, but shift the frequencies of the transverse modes away from the frequency of the lowest order mode, improving the selectivity of cavity modes.  We calculate that for $R = 12$ cm and $\ell_c = 11.9$ cm, the cavity has a resonance at the cesium hyperfine transition frequency (9.2 GHz), its free spectral range (FSR) is $c/2 \ell_c = 1.26$ GHz, the beam radius at the waist is 2.50 cm, the beam radius at the reflectors is 3.53 cm, and the transverse mode spacing is 0.2487 times the FSR, or about 313 MHz.

We estimate the field amplitude at the interaction region as follows.  We choose the spacing between the parallel plates of the transmission line to be 1 cm, and the conductor width 7.5 cm.  These dimensions yield a characteristic impedance of the transmission line of 50 $\Omega$, and allow for a reasonable clearance of the atomic beam in the space between the conductors.  With a copper thickness on the reflectors of 170 nm, we calculate a reflection coefficient of 0.9992.  Note that this thickness is smaller than the skin depth $\delta $ = 680 nm of copper at this frequency, so the transmission losses are small, but not negligible.  With this reflectivity, the cavity losses due to reflection are of the same magnitude as the losses $L$ due to other mechanisms, primarily conduction losses in the upper and lower conducting plates, and diffraction losses due to the finite size of the conductor.  (These results come from our numerical analysis of the cavity modes, which we discuss next.)  For an r.f.\ input power of 250 mW incident on the cavity from either side, the incident voltage amplitude is 5.0 V, and the electric field of the traveling wave incident on the cavity is $\varepsilon_{\rm in}^+$ = 5.0 V/cm.  The amplitude of the traveling wave inside the cavity is 
\begin{displaymath}
   \varepsilon^+ = 2 \ \frac{t \ \varepsilon_{\rm in}^+}{1-r^2 (1 - L)} = 125 \ {\rm V/cm},
\end{displaymath}
where we use $t$ = 0.04 for the transmission coefficient of the reflector and $(1-L) = r^2$.  The factor 2 results from symmetric inputs from the two sides.  At the anti-node of the field, where the amplitudes of the two traveling waves inside the cavity add in phase, the field amplitude is twice this value, or 250 V/cm.  This is the value of the r.f.\ field amplitude that we used in Sec.~\ref{sec:SignalSize} to estimate the signal size.  In making this estimate, we have not included the absorption of the copper reflector, which reduces the amplitude, or the increase of the wave amplitude as the wave propagates to the waist of the Gaussian profile.

\begin{figure*}
\vspace{-0.2in}
  \begin{center}
  \includegraphics[width=4.0cm]{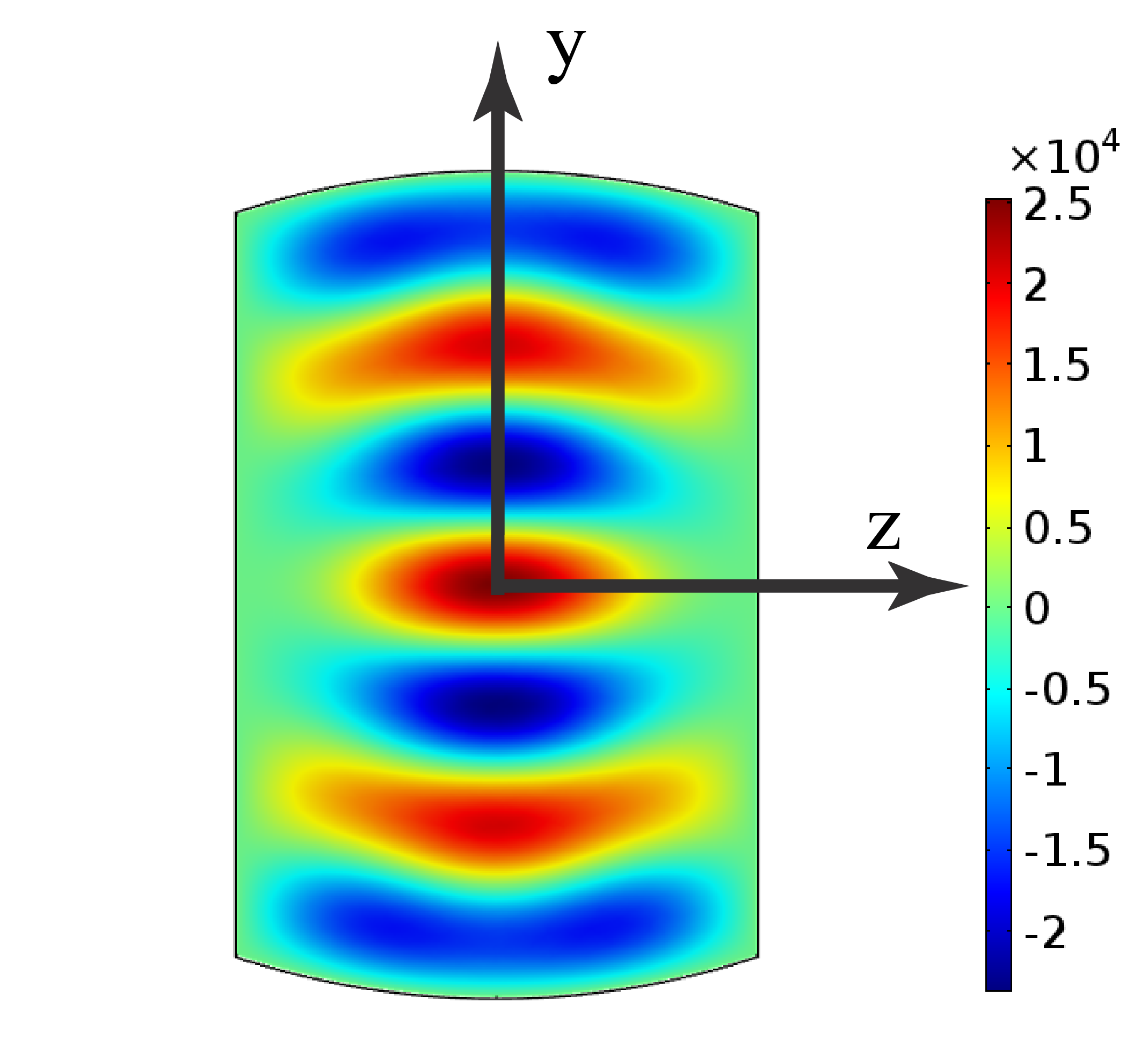} \hspace{0.2in}
    \includegraphics[width=4.0cm]{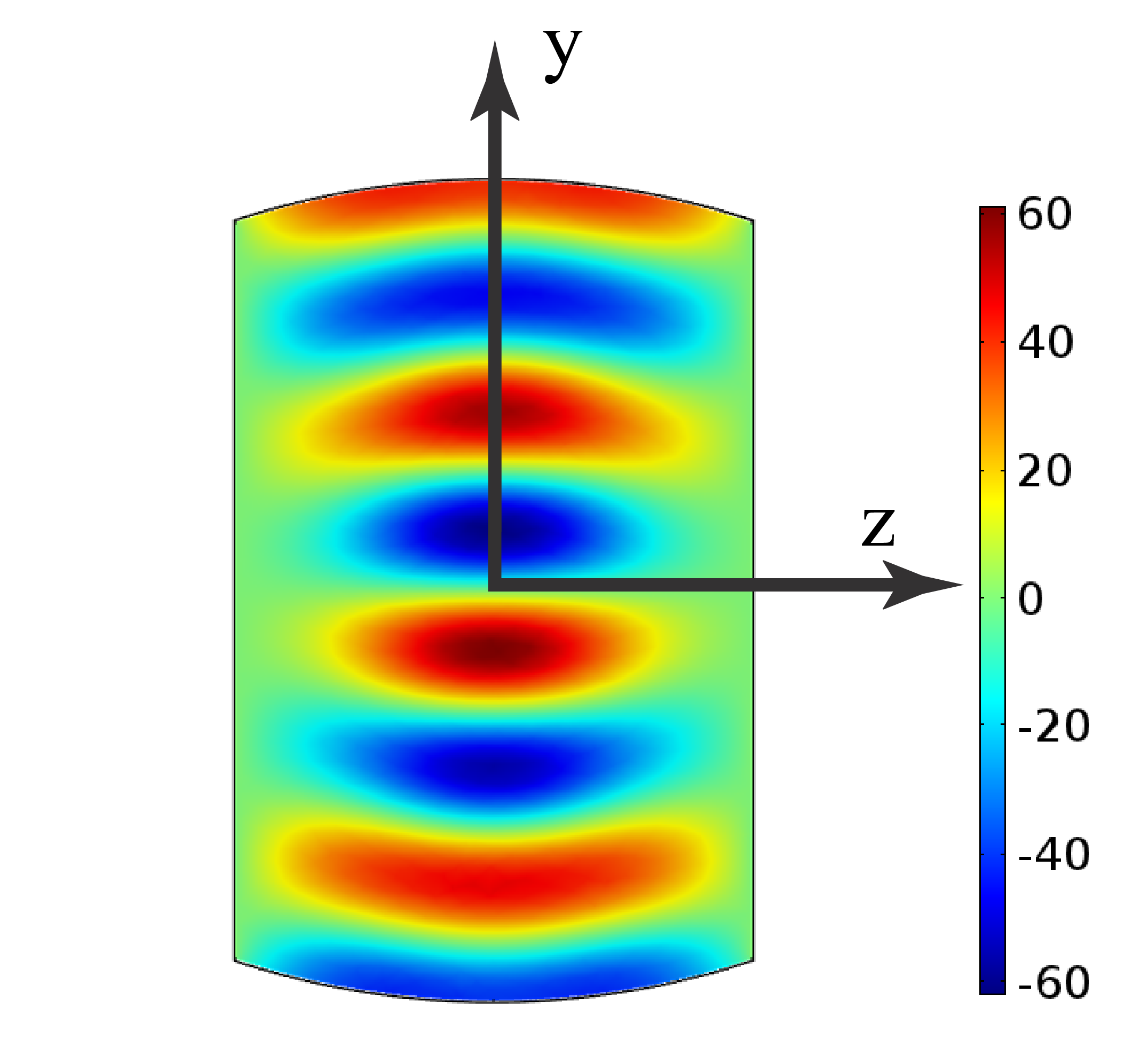} \hspace{0.2in}
       \includegraphics[width=4.0cm]{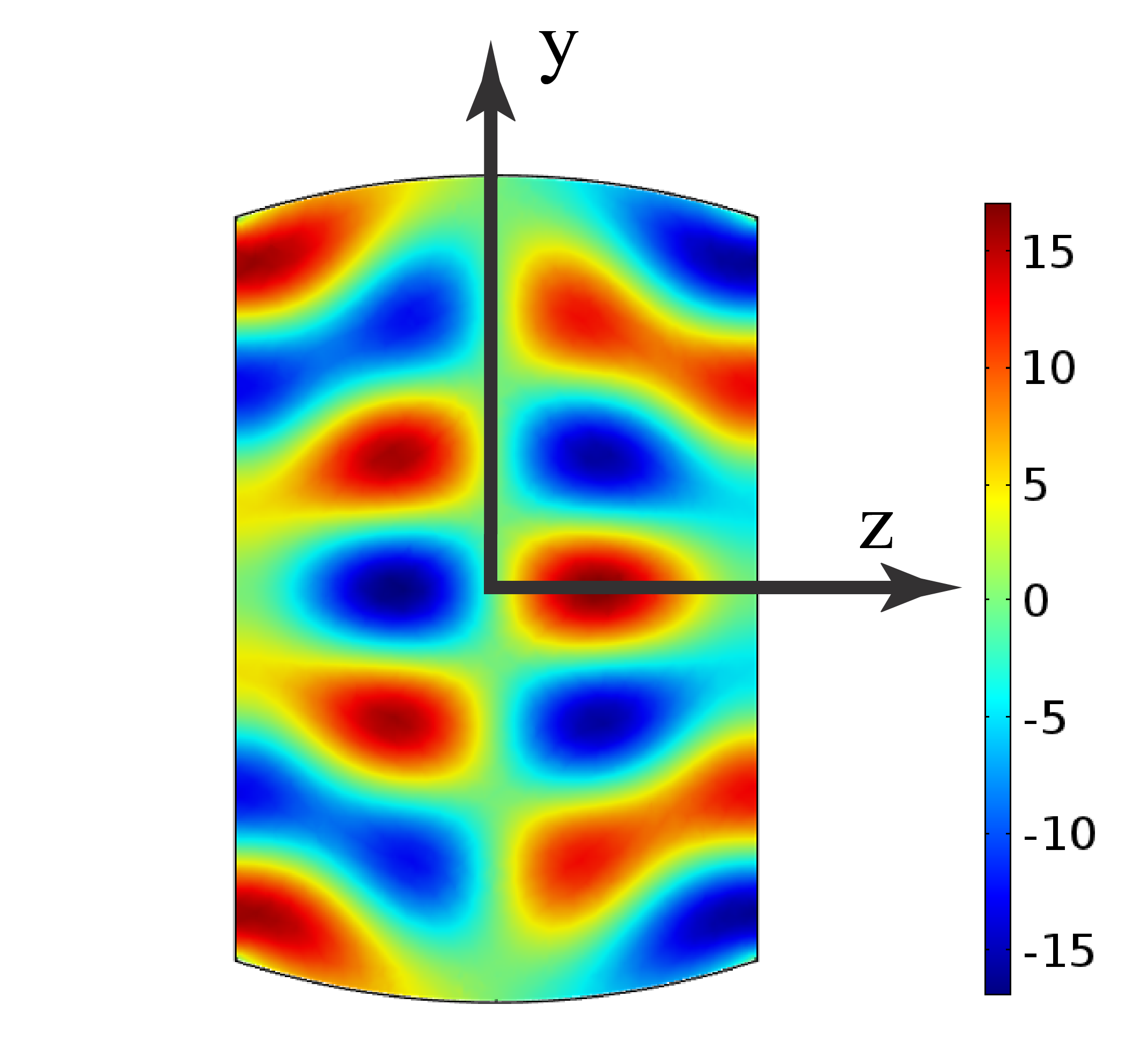} \\ (a) \hspace{1.6in} (b)\hspace{1.65in} (c) \hspace{0.4in}
 \caption{
  (Color online) Color maps of (a) $Re [\varepsilon_x^{\rm rf}(y,z)]$, (b) $Im [h_z^{\rm rf}(y,z)]$, and (c) $Im [h_y^{\rm rf}(y,z)]$ for the lowest-order mode supported by the r.f.\ cavity.  Units for $\varepsilon^{\rm rf}$ are V/m, for $h^{\rm rf}$ are A/m.  } 
  \label{fig:mode_patterns}
\end{center} 
\end{figure*}

\begin{figure}[b!]
\vspace{-0.2in}
  \begin{center}
  \includegraphics[width=7.0cm]{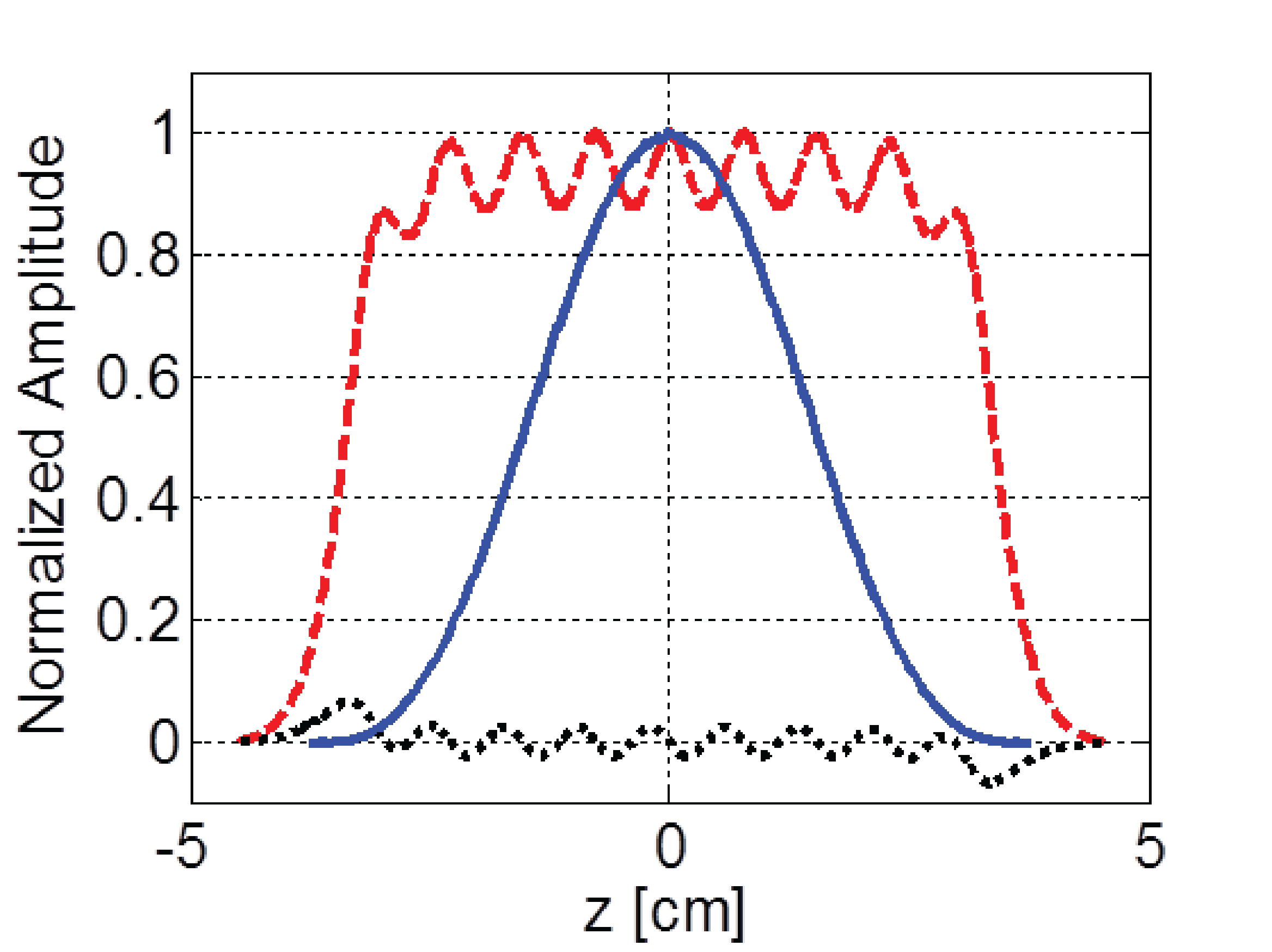} 
 \caption{
  (Color online) Field profiles of $E_z^0(z)$ (red dashed) and $E_x^0(z)$ (black dotted), each normalized to the maximum magnitude of $E_z^0(z)$, at the center of the region between the parallel conducting plates.  Also shown is the Gaussian r.f.\ field amplitude, $ \varepsilon_x^{\rm rf}(0,z)$ (blue solid).    } 
  \label{fig:Field_Profiles}
\end{center} 
\end{figure}

In order to determine more-detailed field parameters, we have carried out a series of numerical simulations of the cavity mode using Comsol MultiPhysics.  These simulations allow us to determine the effects of resistive losses of the parallel plates, the thickness of the reflective copper layers, and the finite width of the cavity on the cavity $Q$; the effect of the gaps in the conductor between the d.c.\ bias pads; and the uniformity of the static electric field in the interaction region.  We show the three primary components, $Re [\varepsilon_x^{\rm rf}(y,z)]$, $Im [h_z^{\rm rf}(y,z)]$, and $Im [h_y^{\rm rf}(y,z)]$, of the simulated r.f. field mode in Fig.~\ref{fig:mode_patterns}.  We note very close agreement of the components $\varepsilon_x^{\rm rf}(y,z)$ with the analytic result in Eq.~(\ref{eq:exSWfoc}) and $h_z^{\rm rf}(y,z)$ with Eq.~(\ref{eq:hzSWfoc}).  
The component $Im [h_y^{\rm rf}(y,z)]$ would be negligible for a weakly focused beam, but since in our geometry, $w_{\rm rf} \sim \lambda$, this component survives.
For this figure, the separation between the upper and lower conducting planes of the PPTL and the width of the conductors are as before, 1.0 cm and 7.5 cm, respectively, as are the radius of curvature of the cylindrical reflectors $R$ = 12.0 cm, and the reflector separation $\ell_c$ = 11.9 cm.  With the thickness of the copper reflector layers equal to 200 nm, we determine a cavity $Q$ of $9000$, while for a 1.5 $\mu$m $\sim 2 \delta$ layer, the $Q$ increases to 13,000.  In the latter case, the $Q$ is limited primarily by the resistive losses in the conductors and diffraction losses of the finite width of the reflectors.    
For a cavity $Q$ of 9000, the linewidth of the transmission peak of the cavity is $\Delta \nu = \nu_0 / Q \sim $ 1 MHz.  We show the computed Gaussian r.f.\ field amplitude, $ \varepsilon_x^{\rm rf}(0,z)$ across the interaction region as the solid blue curve in Fig.~\ref{fig:Field_Profiles}.
The diameter of the cavity mode agrees well with $2w_0$ = 5.0 cm that we determined analytically earlier.  

We used the Eigenfrequency module and frequency domain analysis to carry out these calculations, and determined the quality factor of the cavity as the ratio of the energy stored inside the cavity to the diffraction and dissipation losses.  We obtained the field patterns by launching a 9.2 GHz plane-wave-like electric field on the parallel plate transmission line towards the cavity, exciting a TE$_{q,n}$ cavity mode, where indices $q$ and $n$ label the transverse and longitudinal modes.  The mode spacing between the TE$_{q,n}$ and TE$_{q,n+1}$ mode agrees well with the 1.26 GHz FSR that we determined earlier.  We used a trial-and-error approach to reduce the diffraction losses by varying the cavity parameters, such as the width and height of the cavity, while maintaining the resonant mode frequency close to 9.2 GHz.   

In order to calculate the r.f.\ field distributions in a more refined manner in the interaction region, we added about ten thousand times more mesh points in the vicinity of the interaction region.  Higher mesh point density helped to reduce errors that are present in the interpolation schemes, without compromising the eigenfrequency calculations.  We used ten bias pads, with the spacing between the pads about one tenth the width of the pads. As long as the transmission lines are thin (less than 0.1 mm), the gaps have little impact on the r.f.\ fields.
We found that neither horizontal nor vertical misalignment of the cylindrical reflectors affects the field patterns or the $Q$ factor, for misalignment less than 1 degree.  

We have also modeled all components of the static electric field $\mathbf{E}^0$, and shown that with an array of 10 bias pads and $\Delta V$ = 100 V between pads, we can generate a relatively uniform field of magnitude $E_z^0(z) \sim$140 V/cm.  We show this field, normalized to its maximum value, as the red dashed line in Fig.~\ref{fig:Field_Profiles}.  We also show $E_x^0(z)$ in the plot (black dotted line), which is small in magnitude, and has an average value of zero.  The non-uniform part of $E_z^0(z)$, seen in Fig.~\ref{fig:Field_Profiles} as a nearly sinusoidal modulation of amplitude $\sim$7\% of the constant part, has little impact on the measurement.  We can see this by integrating the product $E_z^0(z) \varepsilon_x^{\rm rf}(z)$ across the interaction region in $z$.  For the case of ten bias pads, as shown, the correction to the signal due to the sinusoidal modulation is less the 0.7\% of the signal.  
We can also see in this figure that the width of the Gaussian shaped r.f.\ field profile is somewhat less than the width of the d.c.\ field, allowing us to avoid fringe effects of the d.c.\ field near the edges of the conductors.

We have used these simulations of the field amplitudes, and their variation through the interaction region, to estimate systematic contributions to the PNC signal.  We discuss these contributions in the next section.

\section{Estimation of magnetic dipole contributions to the PNC signal}\label{sec:systerrors}
%The r.f.\ and laser field amplitudes in the measurement will be time independent, but the motion of the atoms across the interaction region introduces an effective time variation.  The spatial extent of the r.f.\ field is relatively larger than that of the Raman beam.  
In this section, we will make use of the field simulations of the previous section in order to estimate the expected systematic contributions to the PNC signal.  The primary contributions that must be considered are the magnetic dipole terms, due to the relatively large magnetic dipole moment $M$ on this transition.  As we have shown, the primary magnetic field components of the r.f.\ field are $h_z^{\prime \prime}(y,z)$ and $h_y^{\prime \prime}(y,z)$, where we use primed (double-primed) variables for the real (imaginary) part of the field quantities, and omit the superscript `rf'.  By setting up the geometry of the experiment to make the atomic beam cross the r.f.\ field at the center of the cavity, where the component $h_z^{\prime \prime}(y)$ is minimal, the magnetic dipole contributions to the signal from any individual atom can be reduced.  Furthermore, the contributions from atoms on one side of the node are of opposite sign to those on the other side of the node, and the net magnetic dipole contribution can be suppressed even further.  In this section, we use the numerical simulations of the fields supported by the resonant cavity to explore the magnitude of magnetic dipole contributions to the PNC signal.  

The net contribution of the $h_z^{\prime \prime}(y,z)$ term can be minimized by adjusting the relative position $\Delta y$ of the center of the atomic beam relative to the node of the magnetic field.  (No control of the $x$-position is necessary, since the fields are uniform in this direction.)  For a uniform beam density with beam radius $b$, the magnitude of the $h_z^{\prime \prime}(y,z)$ contribution grows as $k \Delta y$ (independent of $b$).  With piezoelectric adjustment and feedback control of the distance $\Delta y$, such that $\Delta y$ is maintained to a value less than, say, 10 $\mu$m, then the magnetic dipole signal is suppressed by an additional factor of $2 \times 10^{-3}$.  This approach to suppression of this magnetic dipole contribution is applicable even with less than ideal beam symmetries, such as an asymmetric beam density, a misalignment of the beam with the z-axis, or divergence of the atomic beam.  Due to the sign reversal of $h_z^{\prime \prime}(y,z)$ across the node, there must always exist a beam location for which this signal averaged over all the atoms in the beam vanishes.

For the $h_y^{\prime \prime}(y,z)$ contributions, the zero is also guaranteed, but for a different reason.  As can be seen in Fig.~\ref{fig:mode_patterns}, this field component is maximal along the atom beam at the center of the cavity, $y=0$, but is an odd function of $z$.  Since the magnetic dipole contribution scales as the integral of $h_y^{\prime \prime}(y,z)$ across the interaction region, and since $h_y(y,z)$, by Faraday's Law, is proportional to $\partial \varepsilon_x/\partial z$, the path integral of $h_y^{\prime \prime}(y,z)$ across the interaction region depends only on $\varepsilon_x$ at the beginning and end of the path.  But $\varepsilon_x$ is zero far from the center of the PPTL, so this contribution also vanishes.  

For these various reasons, we find that the average value of $\int h_i(z) dz$ (including effects of the divergence of the atomic beam) is equal to zero for all components of $h_i$, where the average is computed over all atoms in the beam, and presuming that we have successfully adjusted the node of $h_z$ to be co-located with the center of the atomic beam.  $\Theta_{\rm M}$ for individual atoms may have non-zero values, but when averaged over all atoms, every one of these terms vanishes.  In Table~\ref{table:GS_systematics}, we have listed the magnetic dipole contributions that appeared in Eq.~(\ref{eq:AMDelmpm1}).  
\begin{table*}
\begin{center}
 \begin{tabular}{|c|c|c|c|c|}
   \hline
   \rule[-0.2cm]{0cm}{0.7cm}Comp.  &  $\overline{\int h_i(z) dz}$ & $\left[ \int h_i(z) dz \right]_{\rm rms}$  &   Magnetic Dipole Contribution   & $[\Theta_{\rm M}]_{\rm rms} / \sqrt{N}$  \\ \hline \hline 
   %$\langle h_i \rangle$ & $\sqrt{\langle h_i^2 \rangle}/\sqrt{N} $ & est. rel. mag. 
 
 \multicolumn{5}{|l| }{Magnetic dipole  contributions in phase with $\mathcal{E}_{\rm PNC}$ } \\ \hline
   $ h_x^{\prime \prime}$     &  0   &  40 $\mu$A &  $\eta_0 M  \left[ \int h_x^{\prime \prime}(z) dz \right]_{\rm rms} / \hbar v \sqrt{N}$ & $8 \times 10^{-9}$     \\ 
   
   $ h_y^{\prime }       $     &  0   &  7 nA & $\eta_0 M  \left[ \int h_y^{\prime }(z) dz \right]_{\rm rms} / \hbar v \sqrt{N}$ & $2 \times 10^{-12}$     \\ 
   
   $ h_z^{\prime \prime}       $     &  0   &  0.1 A & $\eta_0 M  \left[ \int h_z^{\prime \prime}(z) dz \right]_{\rm rms} / \hbar v \sqrt{N} \times (B_x^0/B_z^0)$ & $2 \times 10^{-9}$     \\ 
     
   $ h_z^{   \prime}           $     &  0   &  8 nA & $\eta_0 M  \left[ \int h_z^{\prime }(z) dz \right]_{\rm rms} / \hbar v \sqrt{N} \times (B_y^0/B_z^0)$ & $2 \times 10^{-16}$     \\ \hline
     
 \multicolumn{5}{|l| }{ Magnetic dipole contributions in quadrature with $\mathcal{E}_{\rm PNC}$ } \\ \hline
   
   $ h_x^{    \prime}$         &  0   &  5 nA &  $\eta_0 M  \left[ \int h_x^{    \prime}(z) dz \right]_{\rm rms} / \hbar v \sqrt{N}$ & $1 \times  10^{-12}$     \\ 
   
   $ h_y^{\prime \prime}       $     &  0   & 50 $\mu$A & $\eta_0 M  \left[ \int h_y^{\prime \prime}(z) dz \right]_{\rm rms} / \hbar v \sqrt{N}$ & $1 \times 10^{-8}$     \\ 
   
   $ h_z^{\prime  }           $     &  0   &  8 nA & $\eta_0 M  \left[ \int h_z^{\prime   }(z) dz \right]_{\rm rms} / \hbar v \sqrt{N} \times (B_x^0/B_z^0)$ & $2 \times 10^{-16}$     \\ 
     
   $ h_z^{   \prime \prime}   $     &  0   &  0.1 A & $\eta_0 M  \left[ \int h_z^{\prime \prime}(z) dz \right]_{\rm rms} / \hbar v \sqrt{N} \times (B_y^0/B_z^0)$ & $2 \times 10^{-9}$     \\ \hline
 \end{tabular}
 \end{center}
 \caption{Estimates of potential contributions to the atom signal due to magnetic dipole interactions.  For comparison, the amplitude of the PNC-induced term $|\Theta_{\rm PNC}|$ is $ \mathcal{E}_{\rm PNC} \int \varepsilon_x^{    \prime}(z) dz  / \hbar v$, which we evaluate as $5.6 \times 10^{-6}$.  We have organized these terms by those that add in phase to the $\mathcal{E}_{\rm PNC}$ term, followed by those that add in quadrature to the $\mathcal{E}_{\rm PNC}$ term.   In the second column, we list the average value of field component, averaged over the interaction region, which is zero for each component.  In the third column, we list the r.m.s.\ value of the field component.  In the right column, we list the contribution of this term.  All magnetic dipole contributions are suppressed to less than 0.2\% of the $\mathcal{E}_{\rm PNC}$ term.  } \label{table:GS_systematics}
 \end{table*}
These terms can potentially contribute to $\Delta m = \pm 1$ transitions, and are therefore candidates for obscuring the $\mathcal{E}_{\rm PNC}$ signal.  In the second column of this table, we list the average value of each of the $\int h_i(z) dz$ terms, which we have already argued must vanish in each case.  A better metric for comparison is therefore the standard deviation of the distributions of the relevant path integrals of $h_i(y,z)$, which we list in the third column of Table~\ref{table:GS_systematics}.  For this calculation, we computed the distribution of the integrals $\int h_i(z) dz$ (separately for the real and imaginary parts) over various straight-line paths through the interaction region, and determined the width of this distribution by calculating the root-mean-square value, $[\int h_i(z) dz]_{rms}$.  
%We have presented this value for each component of $h^{\prime}$ and $h^{\prime \prime}$ in the table.  
After $N$ atoms have traveled through the interaction region, the standard deviation of the mean of these integrals is $[\int h_i(z) dz]_{rms}/\sqrt{N}$.  We use $N = 3 \times 10^{12}$ for this purpose, the number of atoms that we computed would be necessary to produce a shot-noise limited measurement with an uncertainty of 3\% in Section~\ref{sec:SignalSize}.  Multiplying by $\mu_0 M /(v \hbar)$ (which numerically is equal to 380 A$^{-1}$, which we compute using $M = e a_0 \alpha/2 = 3.1 \times 10^{-32} $ Cm and $v$ = 270 m/s) yields the rms magnitude of each magnetic dipole term.  For the Zeeman mixing contributions, we use a value of $10^{-4}$ for the fractional transverse magnetic field amplitudes $B_x^0/B_z^0$ and $B_y^0/B_z^0$.  As can be seen in Table~\ref{table:GS_systematics}, the greatest of any of these magnetic dipole terms is of magnitude $1 \times 10^{-8}$.  
For comparison, the PNC term integrated across the interaction region is $ \mathcal{E}_{\rm PNC} \int \varepsilon_x(z) dz/v$, which we compute as $5.6 \times 10^{-6}$ for the same set of field parameters.  Since these rms magnetic dipole contributions are much less than the PNC term in each case, we conclude that the magnetic dipole contributions can be sufficiently reduced, and that the PNC measurement in this geometry can be successfully executed.

Finally, we must consider excitation of the $\Delta m = 0 $ transition, which, if present at sufficient rates, could obscure the signal.  Excitation of this transition is strongly driven by the $h_z$ field component.  This transition is suppressed, however, by the application of $B_z^0$, which produces a Zeeman shift between the magnetic levels.  
For $B_z^0 \sim$ 7 Gauss, the Zeeman shift is $\sim$3 MHz.
When the frequencies of the r.f.\ and Raman beams are tuned to resonance with the $\Delta m = \pm 1 $ transition, the $\Delta m = 0 $ transition is off resonant.  To gauge the degree of this excitation, we must estimate the linewidth of the transition.  
Lifetime broadening of ground states and the collisional linewidth in an atomic beam are negligible.  There is a small Doppler broadening of the transition due to divergence of the atomic beam.  $\Delta \nu_D = \nu_0 \Delta v_t/c$, where $\nu_0$ is the 9.2 GHz transition frequency, and $\Delta v_t$ is the transverse velocity spread of the atomic beam.  Using 1 mm apertures separated by 40 cm to form the atomic beam gives a divergence angle of $\sim$2.5 mrad, and with a mean atomic velocity of 270 m/s, $\Delta v_t \approx $ 0.7 m/s, leading to a Doppler width of $\Delta \nu_D \approx $ 20 Hz.  The broadening due to the finite interaction time of the atoms with the r.f.\ field as they pass through the interaction region is $\sim v / 2 \pi w_{\rm rf}$ = 2 kHz.  This transit time broadening and the linewdith of the r.f.\ source are the primary broadening mechanisms that we expect in this atomic beam geometry.  
Since the linewidth for the transition is much less than the Zeeman shift, excitation of this line is very weak.  Furthermore, when we use phase modulation of the r.f.\ field and phase-sensitive detection of the signal, only the signal resulting from the interference between the Raman transition and the r.f.\ excitation is detected.  The Raman transition on the $\Delta m = 0$ transition is highly suppressed by careful orientation of these polarizations and by the Zeeman shift, and so the net interference by excitation of the $\Delta m = 0$ line is expected to be below detection limits.  

To reduce systematics in this measurement, we have several tools and metrics available.  These include: translation of the r.f.\ cavity about the atomic beam, reversals of $\mathbf{E}^0$, $\mathbf{B}^0$, and the initial projection state ($F$, $m$) in which the atoms are prepared, and application of transverse $B_x^0$ or $B_y^0$ field components to intentionally introduce systematic effects.  We expect that these tests and studies will likely require the bulk of our attention as we carry out these measurements.

\section{Velocity spread of an atomic beam }
In this section we examine two effects of the distribution of velocities in the atomic beam on the signal, as well as the variation of the intensity of the Raman coupling laser beams.

\begin{figure}
  \includegraphics[width=7.5cm]{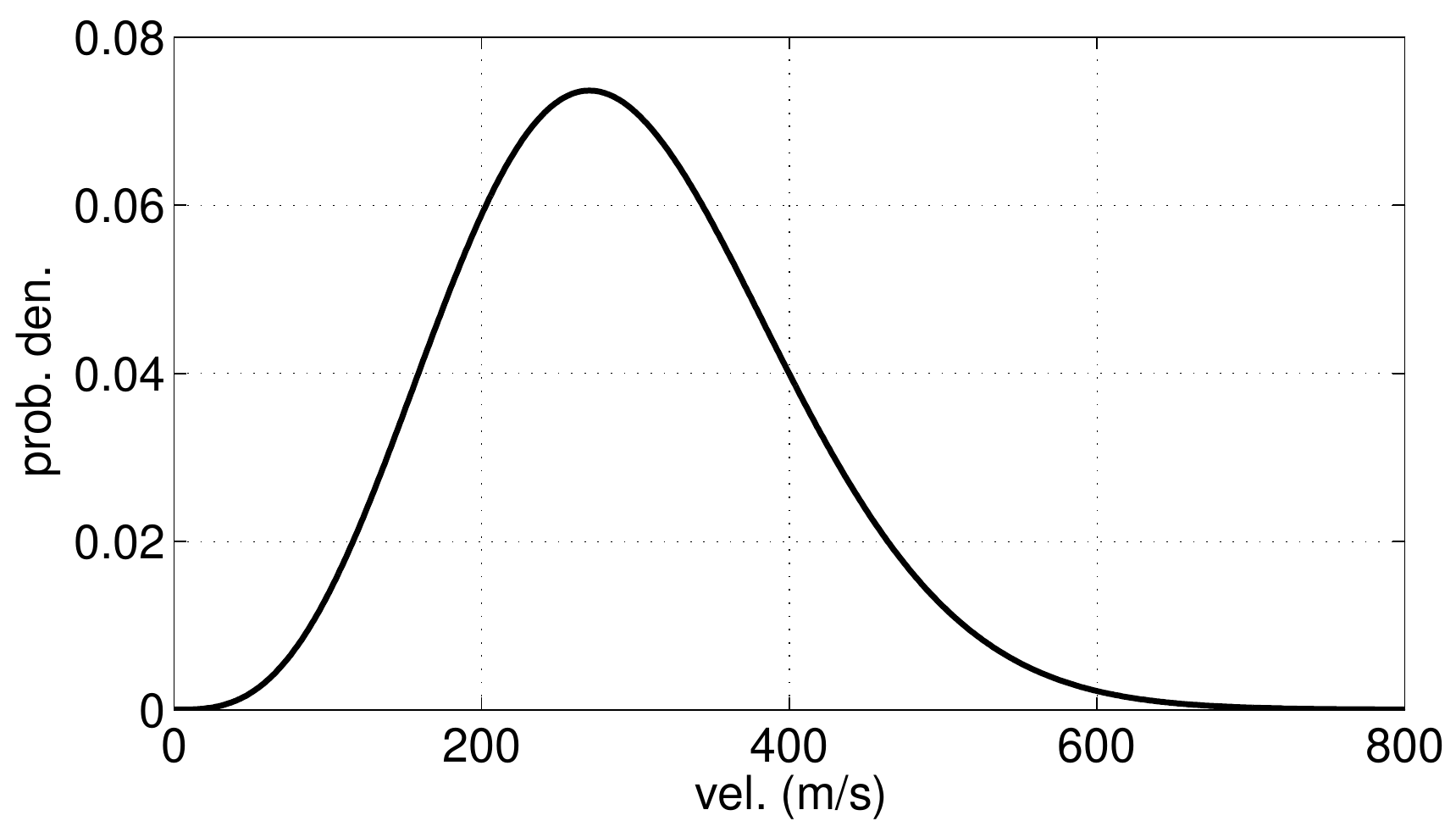}\\
  \caption{(Color online) The computed probability distribution of velocities of atoms in the cesium beam at T = 120$^{\circ}$ C.  The peak velocity is 270 m/s, and the standard deviation is 106 m/s.      }
  \label{fig:vel_dist_at_beam}
\end{figure}

The interaction time for an individual atom depends upon its velocity, and we show the distribution of the atomic velocities in Fig.~\ref{fig:vel_dist_at_beam}.
When the oven temperature is 120 $^{\circ}$C, the peak velocity of the atoms is 270 m/s, and the standard deviation of the distribution is 106 m/s.  
The interaction time will vary among the atoms due to the distribution of their velocities, resulting in a variation of the interaction angle $\Theta$ for each of the interactions, as shown in Eq.~(\ref{eq:Thetaeq1ovintomegadz}).  The interaction affected most significantly is $\Theta_{\rm Ram}$, since an increase or decrease of this term tends to reduce the signal gain factor $\sin(2 | \Theta_{\rm Ram}| ) $ in Eq.~(\ref{eq:cfeqdcpmodterm}).  
We have analyzed this reduction of signal by numerically averaging $\sin(2 | \Theta_{\rm Ram} |) $ over the velocity distribution shown in Fig.~\ref{fig:vel_dist_at_beam}, and find that this reduction is $\sim$23\%.  Since the Stark and PNC amplitudes are affected in the same way, the only effect is a reduction of the signal magnitude; it does not otherwise affect the accuracy of the measurement. 

The distribution of atomic velocities also reduces the signal through the variation in the flight time of atoms from the interaction region to the detection region in the measurement apparatus.  This effect is much less significant in our apparatus, for the parameters of our measurement.  In our system, this distance from the interaction region to the detection region is $\sim$13 cm, so the average time required to reach the detection region is only 0.5 ms.  Due to the width of the velocity distribution, the spread in arrival times is $t \Delta v /v \approx 0.2$ ms.  This satisfies the requirement that this $\Delta t$ is much smaller than the inverse of the modulation frequency modulation of the signal, which in our previous measurements was 150 Hz, and the loss of signal is a few percent.  Once again, this loss only affects the signal size, but does not affect the accuracy of the measurement.

Finally, we examine the effect of the Gaussian intensity distribution of the laser beams that drive the Raman transition in the atoms.  Since for an optimized measurement, the interaction angle $\Theta_{\rm Ram}$ is adjusted to a value of $\pi/4$, if we set the Raman beam amplitudes to this value on the axis, the angle for any off-axis atoms is less, decreasing our signal.  To ensure that all atoms experience the same field amplitude to within 10\% of the peak value, an atom beam diameter of 1 mm requires a beam radius of the Raman beam of 7 mm.  (This dimension is consistent with the value of $w_{\rm Ram}$ that we used in our simulation shown in Fig.~\ref{fig:Cevstime}, since the Raman interaction scales as the product of the laser field amplitudes $\varepsilon_z^{\rm R1}$ and $\varepsilon_x^{\rm R2}$.)

\section{Conclusion}
In this paper, we have reported our design of an experimental configuration for the measurement of the weak-force induced transition moment on the ground-state hyperfine transition of atomic cesium, or any of the alkali metal species, in an atomic beam configuration.  We have introduced an r.f.\ resonant cavity based upon a parallel-plate transmission line structure that allows one to generate the very uniform, well-characterized, high-amplitude r.f.\ and static fields required for this measurement.  This cavity design could be applied to any measurement that requires similar levels of controlled fields.  We have carried out detailed numerical simulations of the various field amplitudes of the standing wave, and used these results to estimate the magnitude of the magnetic dipole contributions to the atomic signal.  The atomic beam geometry is especially well-suited to these purposes, allowing suppression of all magnetic dipole contributions, and we estimate that a precision measurement of $\mathcal{E}_{\rm PNC}$ on the ground state hyperfine transition can be achieved within an integration time of a few tens of seconds.

There are several individuals with whom we have benefited through very useful conversations and exchanges, including S. G. Porsev, L. Orozco, M. Safronova, and A. Derevianko.  
We also acknowledge the contributions of M. Y. Shalaginov, S. Bogdanov, M. Swabey, K. Webb, and the group of D. Peroulis, who were very helpful through their advice with the numerical simulations.

\end{document}